\documentclass[aps,balance,superscriptaddress,floats,showpacs,a4paper]{revtex4-2}

\usepackage[english]{babel}

\usepackage{amsmath,relsize,MnSymbol}
\usepackage{graphicx}
\usepackage{physics}
\usepackage{amsfonts}
\usepackage{natbib}
\usepackage[colorlinks=true, allcolors=blue]{hyperref}

\newcommand{\tcr}{\textcolor{red}}
\renewcommand{\tcr}{}

\newcommand{\bs}{\boldsymbol}
\renewcommand{\vec}{\text{vec}}
\newcommand{\ptheta}{\bs{\theta}}
\usepackage{mathtools}
\DeclarePairedDelimiter\bbra{\llangle}{\rvert}
\DeclarePairedDelimiter\kket{\lvert}{\rrangle}
\DeclarePairedDelimiterX\bbrakket[2]{\llangle}{\rrangle}{#1 \delimsize\vert #2}
\DeclareMathOperator*{\argmin}{arg\,min}
\newcommand{\ceil}[1]{\left\lceil #1 \right\rceil}

\begin{document}
\date{\today}
\title{Faster variational quantum algorithms with quantum kernel-based surrogate models}
\author{Alistair W. R. Smith}
\affiliation{QOLS, Blackett Laboratory, Imperial College London SW7 2AZ, United Kingdom}
\email{Corresponding author: alistair.smith18@imperial.ac.uk}
\author{A. J. Paige}
\affiliation{QOLS, Blackett Laboratory, Imperial College London SW7 2AZ, United Kingdom}
\author{M. S. Kim}
\affiliation{QOLS, Blackett Laboratory, Imperial College London SW7 2AZ, United Kingdom}
\begin{abstract}
We present a new optimization strategy for small-to-intermediate scale variational quantum algorithms \tcr{(VQAs)} on noisy near-term quantum processors which uses a Gaussian process surrogate model equipped with a classically-evaluated quantum kernel. \tcr{VQAs} are typically optimized using gradient-based approaches however these are difficult to implement on current noisy devices, requiring large numbers of objective function evaluations. Our approach shifts this computational burden onto the classical optimizer component of these hybrid algorithms, greatly reducing the number of quantum circuit evaluations required from the quantum processor. We focus on the variational quantum eigensolver (VQE) algorithm and demonstrate numerically that these surrogate models are particularly well suited to the algorithm's objective function. Next, we apply these models to both noiseless and noisy VQE simulations and show that they exhibit better performance than widely-used classical kernels in terms of final accuracy and convergence speed. Compared to the typically-used stochastic gradient-descent approach to \tcr{VQAs}, our quantum kernel-based approach is found to consistently achieve significantly higher accuracy while requiring less than an order of magnitude fewer quantum circuit executions. We analyse the performance of the quantum kernel-based models in terms of the kernels' induced feature spaces and explicitly construct their feature maps. Finally, we describe a scheme for approximating the best-performing quantum kernel using a classically-efficient tensor network representation of its input state and so provide a pathway for scaling this strategy to larger systems.
\end{abstract}
\maketitle
\section{Introduction}\label{sec:intro}
Quantum computation (QC) promises an alternative computational framework for solving an array of classically intractable problems. 
Developments in quantum error correction and the engineering of low-noise qubits have allowed recent experiments to push into the domain of fault tolerant single and two qubit operation \citep{Egan2021,Xue2022,Abobeih2022,Postler2022,Noiri2022}. 
These results present us with a roadmap of how scalable fault-tolerant universal quantum computation may be achieved however the exact timescale for this remains uncertain \citep{Preskill2018,Sevilla2020, Preskill_solvay}. 
Until the threshold for fault tolerance is achieved the kinds of algorithms that can be run on-device with appreciable fidelity are limited to circuits of low depth on relatively few qubits.

These noisy intermediate scale quantum (NISQ \cite{Preskill2018}) algorithms provide us with a useful set of benchmarking tools for quantifying the performance of prototype quantum computers while being an interesting field of study in their own right \cite{Bharti2022}. These include algorithms specifically designed to showcase near-term forms of quantum advantage \citep{Arute2019,Pan2020,Wu2021}. While experimental demonstrations of these ``quantum supremacy" tasks are an important step forward in demonstrating the potential of quantum computers, they are not designed to solve useful problems and the extent of their advantage over classical methods is an area of ongoing study \citep{Bulmer2022,Gray2021,Pan2022}. More modest proposals for interesting NISQ applications include variational quantum algorithms (VQAs) and, the often related, quantum machine learning (QML) \citep{Bharti2022,Cerezo2021_vqas,biamonte2017}. 

QML attempts to generalize the methods and results used in classical machine learning to a quantum setting. 
Quantum kernel methods are a promising NISQ-friendly sub-field of QML focused around the generalization of classical kernel-based algorithms. Generally speaking, kernel methods are a broad class of machine learning techniques which use kernel functions to quantify the similarity between data \citep{hofmann2008,shawe2004}. \tcr{A kernel function is a positive (semi-)definite function which quantifies the similarity between its two inputs; it is equivalent to an inner-product between these inputs mapped into a higher-dimensional feature space.} Quantum analogues to classical kernel methods have primarily been applied to the former set of problems, particularly classification tasks \citep{Rebentrost2014,blank2020,park2020,Mengoni2019}, while applications to the latter have been largely unstudied in the literature \cite{Otten2020}.

VQAs are hybrid quantum-classical algorithms which couple the limited computational capabilities of NISQ devices with classical computational resources in an attempt to solve useful tasks \cite{Cerezo2021_vqas}. Typically, a quantum state is prepared with a parameterized ansatz circuit. The parameters of the ansatz are adjusted by a classical optimizer to minimize a cost function calculated from measurements of the state. The ease of implementing low-depth ansatzes and VQAs' resilience against certain types of coherent noise \citep{Fontana2021} make these algorithms a popular choice for demonstrations on near-term devices \citep{kandala2017,Sung2020,self2021}. 

\tcr{While it has been suggested that scaled-up VQAs could provide a route to quantum advantage in the near-term \citep{Farhi2016,Guerreschi2019,Cerezo2021_vqas}, large hurdles exist to achieving this. In particular, barren plateaus can emerge in the optimization landscape due to device noise \cite{wang2021} or the use of overly expressive many-qubit ansatz circuits \citep{Bittel2021,McClean2018,Larocca2021}. These make the optimization in VQAs exponentially expensive preventing them from being scaled to large system sizes. 
Any direct quantum advantage from large-scale VQE would likely require highly problem-specific ansatzes to avoid these barren plateaus and designing such ansatz circuits is an area of ongoing research \cite{grimsley2019,tang2021}.
Despite these hurdles to scalability, VQAs algorithms have emerged as a popular tool for benchmarking current noisy devices.}

There are several key factors to consider when implementing a VQA on a NISQ device. The ansatz circuit must be capable of representing a state which solves the task at hand (to a good approximation) but should not be so expressive that training becomes infeasible. Device noise must also be considered when choosing the ansatz's depth but is typically difficult to fully characterize. Large amounts of noise can deform a VQA’s cost function to the extent that the obtained solution does not solve the original task while compounding trainability issues \cite{wang2021}. This imposes limitations on the depth of ansatz circuits which can be used and has prompted the development of a plethora of noise mitigation strategies \citep{Temme2017,Kandala2019_errmit,Barron2020,Czarnik2021,Funcke2022,Maciejewski2020}. Finally, a classical optimization strategy must be chosen which makes the best use of available data while taking into consideration factors such as noise and its effect on sampling costs. 

In this work we combine concepts from both VQAs and quantum kernel theory to demonstrate how a surrogate model based on classically-evaluated quantum kernels allows noisy VQAs to be solved quickly and with high accuracy. We focus on the variational quantum eigensolver (VQE) algorithm in which the cost function \tcr{is the energy expectation value} for the ansatz's output state with respect to a Hamiltonian of interest. 
\tcr{Our Bayesian strategy only uses quantum circuit executions to query the energy expectation value for ansatz parameters, and not the gradients of this quantity. These points of interest are chosen through classical optimization of a surrogate model built from the observed energy values. By avoiding on-device estimation of energy gradients and employing an easily-optimized surrogate model, our strategy requires fewer quantum circuit executions per-point-queried than gradient based approaches while also converging in far fewer iterations, making it much more sample-efficient in terms of total quantum circuit executions.} Being a global optimization strategy, it is also resilient against local minima.
\tcr{Because our surrogate model is built with a classically-evaluated quantum kernel, which itself is based on the ansatz circuit, our strategy is able to effectively leverage both explicit knowledge about the ansatz circuit with implicit information about the on-device noise processes contained in the observed energy values.}

\tcr{As the optimization strategy outlined here uses direct classical simulations of quantum kernel functions its applicability is limited to relatively narrow and shallow ansatz circuits (up to $\sim20$ qubits) for which classical simulations are feasible. Even at these scales, performing VQE on-device, in order to physically prepare an approximation to a system's ground state is still a highly non-trivial problem due to device noise and the long queues often required for device access.}
To extend our strategy's applicability, we also outline a scheme to produce a classically tractable approximation to the most-suitable quantum kernel for VQE and so provide a framework for applying the optimization strategy to larger systems.

\section{VQE on NISQ devices}
\tcr{The VQE algorithm is a near-term quantum algorithm that prepares approximations of the ground state and ground state energy of a Hamiltonian and has a wide range of potential applications \cite{peruzzo2014, kandala2017}.} This is done by adjusting the parameters of an ansatz circuit to minimize the measured energy of its output state. For Hamiltonian $H$ and an ansatz state $\ket{\psi(\bs{\theta})}=U(\bs{\theta})\ket{0}$ or $\rho(\ptheta)=\ketbra{\psi(\ptheta)}{\psi(\ptheta)}$, \tcr{which is produced by a parameterized unitary evolution $U(\ptheta)$ with a vector of parameters $\ptheta$}, the goal of VQE is to find $\bs{\theta}_{\mathrm{opt}}$ such that
\begin{widetext}
\begin{equation}\label{eq:vqe_goal}
    \bs{\theta}_{\mathrm{opt}}\coloneqq \argmin_{\bs{\theta}} E(\bs{\theta}), \text{ where } E(\bs{\theta})=\bra{\psi(\bs{\theta})}H\ket{\psi(\bs{\theta})}=\Tr{H \rho(\ptheta))},
\end{equation}
\end{widetext}
\tcr{where $\Tr$ is the trace operation.} By the variational principle, the ansatz state that minimizes the energy is an approximation to the ground state and its energy expectation value is an upper bound to the ground state energy. 


The presence of noise on current quantum processors provides a justification for why on-device small system VQE is still an interesting and important task. On-device VQE gives not only \tcr{an approximation to} the desired ground state energy but also an ansatz circuit that prepares a close approximation to the ground state on the noisy device. Obtaining such circuits with classical methods would first require accurate characterisation of the exact noise processes present during the device's operation, i.e. quantum process tomography, which for more than a few qubits is an incredibly resource-intensive task \cite{Mohseni2008}. Being able to physically prepare a close approximation to the ground state of a Hamiltonian is a useful first step for other quantum computing tasks and for benchmarking the device \citep{Lubinski2021,McCaskey2019}. 

Gradient descent-based optimization strategies are ubiquitous in noiseless functional minimization due to their favourable convergence guarantees and are frequently used in VQAs \cite{Cerezo2021_vqas,Sweke2020,Stokes2020qng,grimsley2019,Nakanishi2019,Parrish2019_VQE,parrish2019}. 
The gradient of the energy function with respect to each ansatz parameter can be estimated directly either through finite difference approaches or using parameter shift rules \cite{Wierichs2022}. Both approaches typically require sampling the cost function at two points per parameter (or more if a parameter appears in multiple gates) meaning these parameter-wise gradients are relatively quantum resource-intensive. 
Current devices are sufficiently noisy that only relatively shallow circuits on few qubits produce outputs not dominated by noise. At these scales ansatz-derived barren plateaus are less of a concern however noise-induced flattening of the cost function can still increase the number of samples needed to make accurate gradient estimates \cite{wang2021}.


To avoid the large cost of estimating noisy parameter-wise gradients more sample-efficient and noise-resilient schemes such as the SPSA have been employed in VQAs \citep{kandala2017,Sweke2020,Gacon2021} at the cost of slower convergence. 
This allows approximate gradient descent to be performed more efficiently on noisy objective functions because its update rule requires evaluations at just two points, regardless of how many parameters are being optimized. 
However SPSA is often slow to converge and can become stuck in local minima, requiring multiple restarts, this means a large number of energy evaluations are needed to solve even relatively small VQE problems.
Other optimization strategies avoid gradients altogether and include heuristic schemes \citep{peruzzo2014,Zhu2019,Shen2017} and surrogate model based approaches \cite{Zhu2019,self2021}.

\section{Surrogate models and quantum kernel  functions}\label{sec:qkernels}
An alternative approach to gradient descent-based VQE is to build a surrogate model for the energy function based on previously observed evaluations. This should, ideally, be differentiable such that it can be optimized more effectively than $\tilde{E}(\ptheta)$, the noise-corrupted energy function. This approach forms the core of Bayesian optimization (BO) \cite{snoek2012} in which a surrogate model is trained on observed data and then optimized to choose the most promising location to query the cost function. 
Previous work has shown how Bayesian VQE combined with a novel information-sharing approach can drastically improve the convergence times of a set of related VQE problems \cite{self2021}. 

Typically, a Gaussian process (GP) surrogate model is used for BO and is primarily defined in terms of a kernel function. A kernel function, \tcr{$k:\mathcal{X}\times\mathcal{X}\to\mathbb{R}$}, is a symmetric positive (semi-)definite function of two arguments \tcr{in an input space $\mathcal{X}$} such that $k(\bs{x},\bs{x'})$ quantifies the similarity between $\bs{x}$ and $\bs{x}'$. 
They can also be understood as an inner product in a reproducing kernel Hilbert space (RKHS) such that $k(\bs{x},\bs{x'}) = \bs{\varphi}^T(\bs{x}')\bs{\varphi}(\bs{x})$, where $\tcr{\bs{\varphi}:\mathcal{X}\to\mathcal{F}}$ is a feature map which takes a point $\bs{x}$ in the input space to a point $\bs{\varphi}(\bs{x})$ in the \emph{feature space}\tcr{, $\mathcal{F}$,} induced by the kernel (a vector in the RKHS) \cite{hofmann2008}. 
We give an overview of Gaussian process surrogates and Bayesian optimization in sections \ref{sec:GPs} and \ref{sec:BO}. 

For a GP model to make accurate predictions about a function's value at an unseen point \tcr{its} kernel function should properly quantify the similarity between this point and \tcr{the points where the function value has already been observed}. Ideally, the objective function being modelled should be a linear function in the feature spaced induced by the kernel. If this is true, the \emph{representer theorem} \cite{scholkopf2001} from classical kernel theory shows that the minimizer of any regularized empirical risk functional (the model that is ``optimal" at describing a set of training data) is given by a weighted sum of kernel evaluations with the points in the training data. 

In general the exact functional form of the cost function is unknown, making it difficult to know which kernel function to use. To combat this, flexible kernel functions equipped with hyperparameters are often used. Many of these have the universal approximation property \cite{micchelli2006}, \tcr{meaning an arbitrary well-behaved function at a point $\bs{x}^*$ can be approximated by a finite weighted sum of kernel evaluations between $\bs{x}^*$ and an appropriately sized ``training set" of other points.} To ensure that a GP model equipped with one of these flexible, agnostic kernels yields good predictive accuracy from the available training data the kernel's hyperparameters are typically chosen by maximising the marginal likelihood of the training data. Optimization of a kernel's hyperparameters is often the most computationally expensive stage of fitting a GP model to observed data (see section \ref{sec:GPs}). To date, surrogate model-based approaches to VQAs have made use of these general-purpose hyperparameterized classical kernel functions \citep{self2021,Shaffer2022}, leaving open the question of whether more appropriate kernel functions exist for VQAs.

Schuld \cite{Schuld2021_kernels} has demonstrated that a very general class of quantum models are linear models in the feature space induced by a quantum kernel. A quantum model is described as a quantum circuit consisting of an encoding unitary followed by a (parameterized) measurement. 
If a VQA's cost function has a linear dependence on the measurement outcomes of a quantum circuit then the results in \cite{Schuld2021_kernels} suggest that a quantum kernel based on the same circuit could be used to produce a powerful surrogate model for the VQA's cost function. 

\subsection{Quantum kernel functions}
The two quantum kernels that we will consider in this work are the ``state kernel", which is equivalent to the fidelity between quantum states, and the ``unitary kernel", which quantifies the similarity between two unitary operations in terms of their Hilbert-Schmidt norm. For an ansatz circuit $\ket{\psi(\ptheta)}=U(\ptheta)\ket{\bs{0}}$, with an $n$-qubit input state $\ket{\bs{0}}=\ket{0}^{\otimes n}$, the state kernel evaluated between gate angles $\ptheta$ and $\ptheta'$ is defined as:
\begin{equation}
\begin{split}
k_s(\ptheta,\ptheta')&\coloneqq\abs{\braket{\psi(\ptheta')}{\psi(\ptheta)}}^2\\&=\abs{\bra{\bs{0}}U(\ptheta')U(\ptheta)\ket{\bs{0}}}^2    
\end{split}
\end{equation}
or equivalently by $k_s(\ptheta,\ptheta')\coloneqq\Tr{\rho(\ptheta')\rho(\ptheta)}$ (assuming a pure quantum state). 
As we discuss in section \ref{sec:vqe} and explicitly show in appendix \ref{appendix:feature_spaces}, the energy function is linear in the feature space induced by the state kernel provided this kernel is based on the same circuit used to calculate $E(\ptheta)$. As a result the state kernel is still expected to be effective when building general surrogate models for $E(\ptheta)$. 

The unitary kernel also induces a feature space in which the $E(\ptheta)$ is a linear function. For an ansatz imparting a $d\times d$-dimensional unitary $U(\ptheta)$ the unitary kernel is defined as
\begin{equation}
k_u(\ptheta,\ptheta')\coloneqq\abs{\frac{\Tr{U^\dagger(\ptheta')U(\ptheta)}}{d}}^2,
\end{equation}
which is the (normalized) modulus-squared Hilbert-Schmidt norm between the unitaries $U(\ptheta)$ and $U(\ptheta')$. We show in appendices \ref{appendix:fs_s} and \ref{appendix:fs_u}  that while $E(\ptheta)$ is linear in the feature space induced by this kernel this space typically has a much larger dimension than that of the state kernel and so more data is required to build accurate regression models using it.

We will now explicitly construct the feature maps of these two kernels for an ansatz circuit $\ket{\psi(\bs{\theta})}=U(\bs{\theta})\ket{\bs{0}}$. Note that while a kernel function and corresponding reproducing kernel Hilbert space are unique, the feature map is not unique and is only defined up to an isometry (which preserves the inner product and thereby the kernel). \tcr{We consider $n$-qubit $p$-parameter ansatzes formed of an initial state $\ket{\bs{0}}=\ket{0}^{\otimes n}$ acted upon by $p$ parameterized Pauli rotations, with $n$-qubit Pauli operators $\{P_1,\dots,P_p\}$, which are interleaved with $p+1$ arbitrary fixed unitaries $\{R_1,\dots R_{p+1}\}$. The corresponding unitary $U(\bs{\theta})$ is given by:}
\begin{equation}\label{eq:pauli_ansatz}
U(\bs{\theta})\coloneqq R_{p+1}e^{-iP_p\theta_p/2}R_p\dots R_2e^{-iP_1\theta_1/2}R_1.
\end{equation}
\tcr{``Hardware efficient" ansatzes \cite{kandala2017} admit this description and are formed of individual parameterized Pauli rotations (PPRs) with low-weight Pauli operators; these ansatzes are common in NISQ quantum computing as they often map efficiently onto a device's native gate set \cite{Cerezo2021_vqas}.} \tcr{Ansatzes of the form given by \eqref{eq:pauli_ansatz} are extremely flexible, being able to implement arbitrary parameterized unitaries provided a one uses a suitably many PPRs, a sufficiently sophisticated encoding of the angles $\bs{\theta}$, and a careful choice of the $\{R_i\}$ (although these may require exponentially deep circuits \cite{Barenco1995}).}

We show in appendix \ref{appendix:feature_spaces} that the state kernel's feature space can be constructed in terms of an ansatz-dependent $3^p$-element vector of operators on $\mathcal{H}\otimes\mathcal{H}^*$ (where $\mathcal{H}$ is the qubits' Hilbert space) $\bs{s}$, given by 
\begin{widetext}
\begin{equation}\label{eq:s}
    \bs{s}=\frac{1}{2^p}\left[
    \begin{pmatrix}
    (I \otimes I + P_p \otimes P_p^*)R_p\otimes R_p^*\\
    (I\otimes iP_p^* -iP_p\otimes I)R_p\otimes R_p^*\\
    (I \otimes I - P_p \otimes P_p^*) R_p\otimes R_p^* 
    \end{pmatrix}
    \otimes_K\dots\otimes_K
    \begin{pmatrix}
    (I \otimes I + P_1\otimes P_1^*)R_1\otimes R_1^*\\
    (I\otimes iP_1^* -iP_1\otimes I)R_1\otimes R_1^*\\
    (I\otimes I - P_1 \otimes P_1^*) R_1\otimes R_1^* 
    \end{pmatrix}
    \right]
\end{equation}
\end{widetext}
and an angle-dependent vector $\bs{v}(\bs{\theta})$, given by
\begin{equation}\label{eq:v}
    \bs{v}(\bs{\theta})=\begin{pmatrix}1\\\sin(\theta_p)\\\cos(\theta_p)\end{pmatrix}\otimes_K\dots\otimes_K\begin{pmatrix}1\\\sin(\theta_1)\\\cos(\theta_1)\end{pmatrix}.
\end{equation}
Here we have made a distinction between the tensor products $\otimes$ between the physical and conjugated Hilbert spaces of the qubits and Kronecker products $\otimes_K$ which are used to construct the kernels' feature maps. The $\otimes_K$ act as tensor products between the sub-vectors in $\bs{s}$ and $\bs{v}$ imposing matrix multiplication in the former case and ordinary multiplication in the latter. Note that the elements of $\bs{v}(\ptheta)$ are linearly independent Fourier components and so cannot be decomposed further. 

The state kernel is then
\begin{equation}
    k_s(\bs{\theta},\bs{\theta}') =\bs{v}^T(\bs{\theta}')\bs{S} \bs{v}(\bs{\theta}),
\end{equation}
where $(\bs{S})_{ij}=\bbra{\rho_0}s_i^\dagger s_j\kket{\rho_0}$ is a Hermitian matrix of inner products between the (unnormalized) states $s_i\kket{\rho_0}$ and $s_j\kket{\rho_0}$ ($s_i$ and $s_j$ are the $i^{\mathrm{th}}$ and $j^{\mathrm{th}}$ components of $\bs{s}$, respectively), $\rho_0=\ket{\bs{0}}\bra{\bs{0}}$, and $\kket{X}$ denotes the vectorization of $X$ so that $\kket{\rho_0}=\ket{\bs{0}}\otimes\ket{\bs{0}^*}$~\cite{gilchrist2011}. As $\bs{S}$ is a Hermitian matrix of inner products it is a positive-semidefinite Gram matrix and admits a Cholesky decomposition $\bs{S}=Q^\dagger Q$ (where $Q$ is an upper-triangular matrix). This means that the state kernel can be written as the inner-product of two $3^p$-element vectors, $k_s(\bs{\theta},\bs{\theta}') = \bs{v}^T(\bs{\theta}')Q^\dagger Q \bs{v}(\bs{\theta}) = \bs{\varphi}_s^\dagger(\bs{\theta}') \bs{\varphi}_s(\bs{\theta})$, and so its feature map $\bs{\varphi}_s:\mathcal{X}\to\mathcal{F}_s$ is
\begin{equation}\label{eq:state_fs}
    \bs{\varphi}_s(\ptheta) = Q\bs{v}(\ptheta),\ \text{where}\  Q^\dagger Q=\bs{S}.
\end{equation}

The construction of the feature map for the unitary kernel is similar and involves the same two vectors $\bs{v}(\bs{\theta})$ \eqref{eq:v} and $\bs{s}$ \eqref{eq:s}. The unitary kernel is given by
\begin{equation}
    k_u(\bs{\theta},\bs{\theta}') =\bs{v}^T(\bs{\theta}')\bs{T} \bs{v}(\bs{\theta}),
\end{equation}
where, $(\bs{T})_{ij}={\Tr{s_i^\dagger s_j}}/{4^n}$ is a Hermitian Gram matrix of Hilbert-Schmidt inner products between the operators $s_i$ and $s_j$. It too is positive-semidefinite allowing it to be written in Cholesky form $\bs{T}=B^\dagger B$ (with $B$ upper-triangular), meaning that the unitary kernel is $k_u(\bs{\theta},\bs{\theta}') =\bs{v}^T(\bs{\theta}')B^\dagger B \bs{v}(\bs{\theta})=\bs{\varphi}_u^\dagger (\bs{\theta}')\bs{\varphi}_u (\bs{\theta})$ and its feature map $\bs{\varphi}_u:\mathcal{X}\to\mathcal{F}_u$ is
\begin{equation}\label{eq:unitary_fs}
    \bs{\varphi}_u(\ptheta) = B\bs{v}(\ptheta),\ \text{where}\  B^\dagger B=\bs{T}.
\end{equation}
So far we have assumed that all the gate angles used in the PPRs are independent. In more sophisticated ansatzes whose gate angles are linear combinations of the ansatz parameters one can apply linear transformations to $\bs{v}(\ptheta)$ such that it contains a set of independent Fourier components and so apply a similar analysis. Provided the couplings between gate parameters are linear (for example if multiple parameterized gates share the same parameter), this will act to reduce the dimension of the induced feature space.

\subsection{VQE in a kernel setting}\label{sec:vqe}
For Bayesian VQE to be effective, the kernel used must result in a GP model which can well-approximate $\tilde{E}(\bs{\theta})$, the noisy energy function. An interesting question this poses is whether quantum kernels are more ideally suited to this task than typical classical kernels. To justify that these kernels are well suited, we can analyse noiseless energy landscape $E(\bs{\theta})$ for a Hamiltonian $H$ and an ansatz unitary of the form given in \eqref{eq:pauli_ansatz}. In appendix \ref{appendix:feature_spaces} we show that this is 
\begin{equation}\label{eq:linear_energy}
    E(\bs{\theta})=\bs{h}^T\bs{v}(\bs{\theta}),
\end{equation}
where $h_i=\bbra{R_{p+1}H R_{p+1}^\dagger}s_i\kket{\rho_0}$ and $\bs{s}$ is given in \eqref{eq:s}. Encouragingly, we see that the angle-dependent part of the quantum kernels' feature spaces $\bs{v}(\ptheta)$ makes an appearance, mirroring the decomposition of the energy function in \cite{parrish2019} in terms of a weighted sum of Fourier components. For a suitable set of energy observations, e.g. at the points $\{(\theta_1,\cdots,\theta_p)\in\{0,\pi/2,\pi\}^{\times p}\}$ or $\{(\theta_1,\cdots,\theta_p)\in\{-2\pi/3,0,2\pi/3\}^{\times p}\}$ it is straightforward to see how the components of $\bs{h}$ (the weights of the Fourier components in $E(\ptheta)$) can be directly inferred (the latter set of points giving exactly the result in \cite{parrish2019}).
As the feature spaces of both the state and unitary kernels are related to $\bs{v}(\theta)$ by linear transformations, the energy is a linear function in both these spaces. More explicitly, $E(\bs{\ptheta})=\bs{h}^T Q^{-1}\bs{\varphi}_s(\ptheta) = \bs{h}^T B^{-1}\bs{\varphi}_u(\ptheta)$ ($Q$ and $B$ are defined in \eqref{eq:state_fs} and \eqref{eq:unitary_fs}, respectively and are assumed to be invertible). 

Given a set of training data (pairs of ansatz parameters and corresponding noiseless energy evaluations), the representer theorem \cite{scholkopf2001} guarantees that the minimizer of any regularized empirical loss functional is given by a finite linear sum of weighted kernel evaluations evaluated at the training points (ansatz parameters whose energy values have been observed) \cite{scholkopf2001}. For an $l_2$-regularized least-squares loss the minimizer is kernel ridge regression \cite{williams2006}, which is mathematically equivalent to the posterior mean of a Gaussian process \eqref{eq:gp_mean}. Because these kernels have finite dimensional feature spaces, kernel ridge regression or GP models using them should be able to achieve perfect global prediction accuracy (can perfectly predict $E(\ptheta)$ for all $\ptheta$) if the size of the training data set is greater than or equal to the feature space dimension (assuming the data is linearly independent in the feature space). 

We note that linearity of $E(\ptheta)$ in these kernels' induced feature spaces is achieved without any hyperparameters. This removes the need for any maximum likelihood hyperparameter optimization. 
It also means that the state or unitary kernel between any two observed points (used to construct the Gram matrix $\mathbf{K}$) only needs to be evaluated a single time.

\subsection{Gaussian processes with quantum kernels}
While both quantum kernels look promising for VQE, one might suspect the state kernel to be more suitable as VQE directly concerns the energy of the state produced by the ansatz, rather than its unitary.

\begin{figure*}
    \centering
    \includegraphics[scale=0.6]{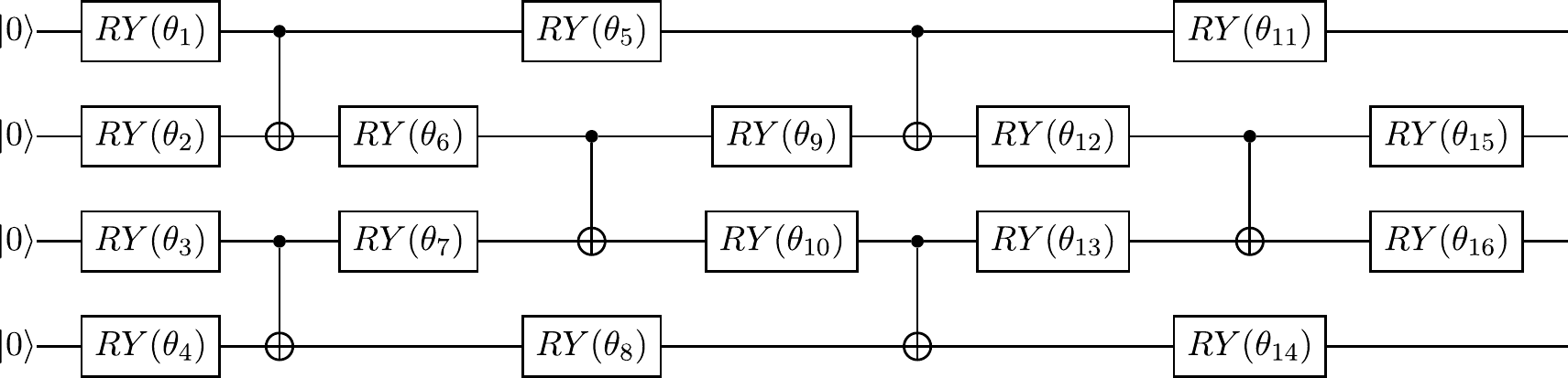}
    \caption{\textbf{Ansatz used for simulations.} This non-periodic ansatz yields a real wavefunction as its output state. The single-qubit gate $RY(\theta_j)$ imparts a unitary $RY(\theta_j)=e^{-i\sigma_y\theta_j/2}$. As the ansatz has 4 layers of $CX$ gates on alternating pairs of adjacent qubits we describe this circuit as being depth 4.}
    \label{fig:ansatz}
\end{figure*}

To evaluate the effectiveness of these quantum kernels versus typical classical kernels in kernel-based VQE regression we first consider the problem of attempting to build a GP model for observations of the noiseless $E(\bs{\theta})$. The Hamiltonian we will use throughout is the anti-ferromagnetic 1D transverse field Ising model with a longitudinal field, given by:
\begin{equation}\label{eq:ham}
    H=J\sum_{\langle i,j\rangle}Z_iZ_j +h_x\sum_iX_i+h_z\sum_iZ_i,
\end{equation}
\tcr{where $J$ is a coupling strength, $h_x$ is the transverse field strength, $h_z$ is the longitudinal field strength, $X_i$ and $Z_i$ are the Pauli $X$ and $Z$ operators acting respectively on the $i^\mathrm{th}$ qubit, $Z_iZ_j$ denotes simultaneous Pauli $Z$ operators on the $i^\mathrm{th}$ and $j^\mathrm{th}$ qubits, and the first sum is taken over nearest-neighbour pairs in a 1D spin chain with periodic boundary conditions.} Throughout we use $J=h_z=0.5$ and $h_x=-0.5$. As this is a real-valued Hamiltonian, its eigenstates can also be taken to be real. This means that for VQE we can limit our ansatz circuit to only yielding real wavefunctions. The ansatz we will use is shown in Fig. \ref{fig:ansatz} and is designed to be efficiently implementable on a quantum processor with nearest-neighbour connectivity between qubits (which helpfully mirrors the nearest-neighbour terms in the Hamiltonian). It consists of parameterised rotations $RY(\theta_j)=e^{-i\sigma_y\theta_j/2}$ and fixed controlled-$X$ gates to build up entanglement.
 
 Let $\mathbf{G}[\mu(\cdot),k(\cdot,\cdot, \bs{\alpha})]$ be a Gaussian process model with covariance/kernel function $k$, kernel hyperparameters $\bs{\alpha}$, and mean function $\mu$. Given a set of observed training and validation energies, $\bs{y}_{\mathrm{t}}=\{E(\ptheta_{\mathrm{t}_1}),\dots,E(\ptheta_{\mathrm{t}_{\mathrm{N_t}}})\}$ and $\bs{y}_{\mathrm{v}}=\{E(\ptheta_{\mathrm{v}_1}),\dots,E(\ptheta_{\mathrm{v}_{\mathrm{N_v}}})\}$  at parameters $\mathbf{X}_\mathrm{t}=\{\ptheta_{\mathrm{t}_1},\dots,\ptheta_{\mathrm{t}_{\mathrm{N_t}}}\}$ and $\mathbf{X}_\mathrm{v}=\{\ptheta_{\mathrm{v}_1},\dots,\ptheta_{\mathrm{v}_{\mathrm{N_v}}}\}$ respectively, we define validation score $R^2_\mathrm{v}$ as
\begin{equation}\label{eq:val_score}
    R_\mathrm{v}^2(\mathbf{X}_\mathrm{v},\bs{y}_\mathrm{v}|\mathbf{X}_\mathrm{t},\bs{y}_\mathrm{t})\coloneqq1-{\frac{\sum_{\bs{\theta}\in \mathbf{X}_\mathrm{v}}[E(\bs{\theta})-\hat{y}_{\mathbf{G}}(\bs{\theta})]^2}{\mathrm{Var}[\bs{y}_\mathrm{v}]}},
\end{equation}
where $\hat{y}_{\mathbf{G}}(\bs{\theta}) = \mathbb{E}[y^*|\ptheta,\mathbf{X}_t,\bs{y}_t]$ is the prediction of model $\mathbf{G}$ at point $\bs{\theta}$ trained with data $(\mathbf{X}_\mathrm{t},\bs{y}_{\mathrm{t}})$ (which, if appropriate, includes hyperparameter optimization). A validation score of $R^2_{\mathrm{v}}=0$ indicates that the GP model is performing as well (with respect to the $L_2$ loss) as a model that always predicts $\mathbb{E}[\bs{y}_\mathrm{v}]$, while a score of $R^2_{\mathrm{v}}=1$ shows perfect prediction across the unseen validation set.

\begin{figure*}
    \centering
    \includegraphics[width=0.5\linewidth]{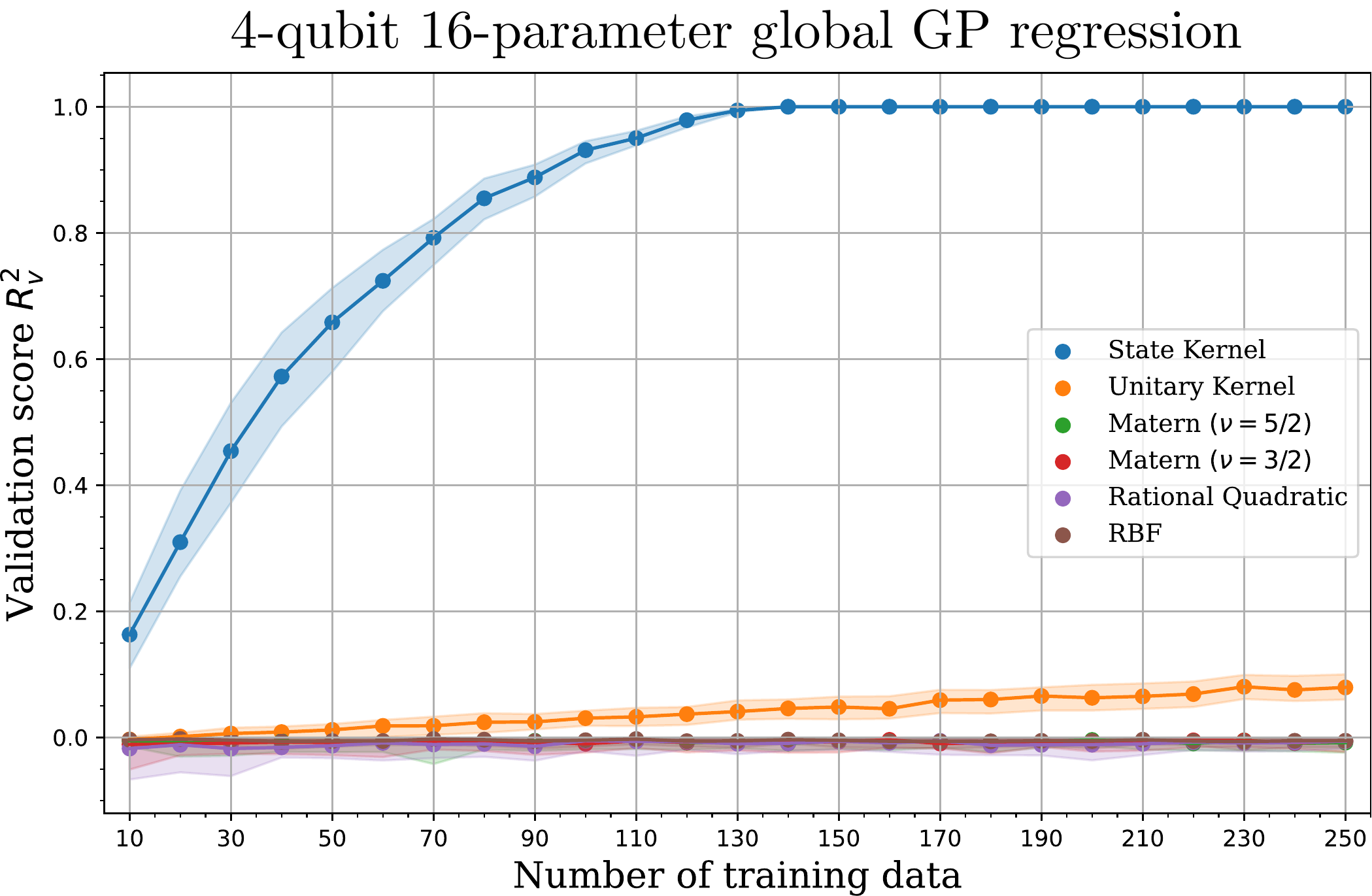}
    \caption{\textbf{Validation score  $R^2_{\mathrm{v}}$ for GP regression models with different kernels learning the energy landscape $E(\bs{\theta})$ for the ansatz given in Fig. \ref{fig:ansatz} and Hamiltonian in \eqref{eq:ham} with $J=h_z=0.5$ and $h_x=-0.5$.} Horizontal axis is the size of the training data set $N_{\mathrm{t}}$, vertical axis is the validation score (defined in \eqref{eq:val_score}). Shown is the median validation scores over 100 repeats (solid lines/dots) and the inter-quartile range of this data (filled). The training and validation points ($\mathbf{X}_{\mathrm{t}}$ and $\mathbf{X}_{\mathrm{t}}$) are chosen uniformly at random (for each of the 16 parameters) in the range $[-\pi,\pi]$ and the corresponding energies $\bs{y}_{\mathrm{t}}$ and $\bs{y}_{\mathrm{v}}$ are evaluated noiselessly on a classical computer. Maximum likelihood hyperparameter optimization with respect to the training data was used to optimize any kernel hyperparameters.}
    \label{fig:gp_full}
\end{figure*}
Fig. \ref{fig:gp_full} shows the improvement in validation score as the size of the training data is increased for GP regression models equipped with different quantum and typical classical kernels. The models are presented with noiseless evaluations of $E(\bs{\theta})$ for the Hamiltonian given in \eqref{eq:ham} ($J=h_z=0.5$ and $h_x=-0.5$). The training and validation sets ($\mathbf{X}_{\mathrm{t}}$ and $\mathbf{X}_{\mathrm{v}}$) are drawn uniformly at random from the interval $\{-\pi,\pi\}$ (for each of the 16 parameters). Both the median $R^2_{\mathrm{v}}$ over 100 repeats (solid line) and the inter-quartile range of the data are shown. The quantum kernel and energy evaluations were performed noiselessly on a classical computer. In all cases, a small diagonal offset was added to the Gram matrices used in GP model prediction $\mathbf{K}\to\mathbf{K}+10^{-10} I$ (equivalent to $\sigma_n^2=10^{-10}$ in \eqref{eq:gp_mean}) to ensure numerical stability.

We see that a Gaussian process model equipped with the state kernel greatly outperforms the other kernels and reaches the optimal validation score of $R^2_{\mathrm{v}}=1$ once the size of the training set reaches $N_{\mathrm{t}}=136$. The validation score of the unitary kernel improves much more slowly, and for $N_{\mathrm{t}}=136$ achieves an $R^2_{\mathrm{v}}<0.1$. However this is still much higher than the scores for the various classical kernels which perform poorly at this task. These classical kernels yield validation scores $R_\mathrm{v}<0$ and show negligible signs of improvement as the training data set grows. 

It is clear, and perhaps unsurprising, that a state kernel built from the same circuit used for a VQE problem provides a much more accurate similarity measure between data points than the other kernels we consider. \tcr{This is because the energy function is linear in the finite-dimensional feature space induced by the state kernel (as discussed in section \ref{sec:vqe}). While $E(\ptheta)$ is also linear in the unitary kernel's feature space, we show in section \ref{sec:feat_construction} that this kernel's induced feature space is typically quadratically larger than that of the state kernel. To build a globally accurate GP model, the observations must span a significant portion of this space which explains why $k_s$ vastly outperforms $k_u$ at this task. The classical kernels we use are universal but do not appear particularly well-suited to the complicated periodic and highly oscillatory $E(\ptheta)$. As this is $16$-parameter problem, the density of the couple of hundred observations in the input parameter space will be far lower than what is required for these classical kernels to make good use of their universal property.}  

For Bayesian optimization we are not always interested in creating an accurate global (i.e. $\theta_i\in[-\pi,\pi]$) regression model as in Fig. \ref{fig:gp_full}. While good global accuracy is helpful, in Bayesian optimization only promising regions of parameter space (those that look close to the desired extremum) are explored in detail. On smaller parameter scales the energy landscape will be less dramatically oscillatory and the density of observed data will be higher meaning that the classical kernels may have a better chance of accurately interpolating between observations.

To test this we examine the validation score for GP models when performing local regression of a VQE energy landscape, restricted to a smaller region of the parameter space. In this case, the training set $\mathbf{X}_\mathrm{t}$ and validation set $\mathbf{X}_\mathrm{v}$ are generated by first picking a common anchor point $\bs{\theta}^{(0)}$ whose elements are chosen uniformly at random in the range $[-\pi,\pi]$. The circuit parameters for $\mathbf{X}_{\mathrm{t}}$ and $\mathbf{X}_{\mathrm{v}}$ are then obtained by sampling parameter vectors $\bs{\theta}$ with elements drawn uniformly at random from the interval $\theta_i\in[\theta^{(0)}_i-\pi/s,\theta^{(0)}_i+\pi/s]$, where $s$ is a scale reduction factor.

\begin{figure*}
    \centering
    \includegraphics[width=\linewidth]{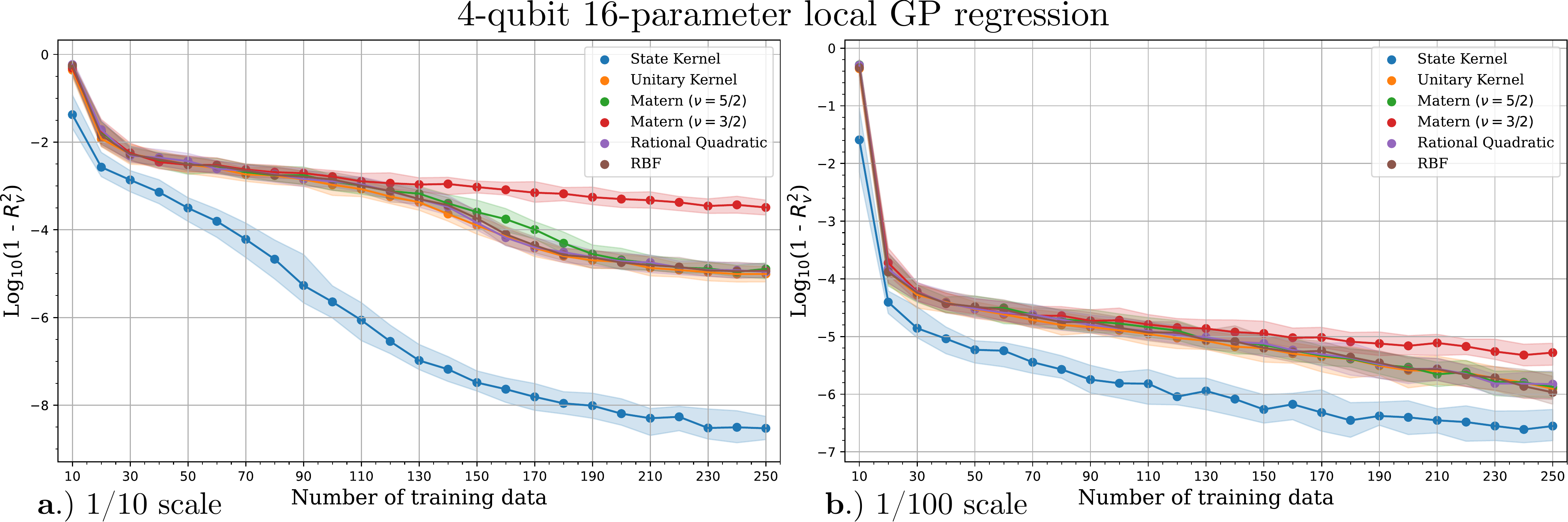}
    \caption{\textbf{Log prediction error $\log_{10}(1-R^2_{\mathrm{v}})$ for GP regression models with different kernels learning the energy landscape $E(\bs{\theta})$ over reduced scales for the same ansatz and Hamiltonian as in Figure \ref{fig:gp_full}.} Horizontal axes are the size of the training data set $N_{\mathrm{t}}$, vertical axes give the logarithm of the prediction error $\log_{10}(1-R^2_{\mathrm{v}})$. Shown are the median over 100 repeats (solid lines/dots) and the inter-quartile range of the data (filled). For each repeat, the GP models are trained and their predictions validated using energy evaluations at ansatz parameters sampled around a randomly chosen anchor point $\bs{\theta}^{(0)}$. To generate the training and validation points ($\mathbf{X}_{\mathrm{t}}$ and $\mathbf{X}_{\mathrm{v}}$), each gate angle, $\theta_i$, is sampled uniformly at random with $\theta_i\in[\theta_i^{(0)}-\pi/s,\theta_i^{(0)}+\pi/s]$, where $s$ is a scale reduction parameter. The corresponding energies $\bs{y}_\mathrm{t}$ and $\bs{y}_\mathrm{v}$ and evaluated noiselessly on a classical computer. \textbf{a.)} $s=10$. \textbf{b.)} $s=100$. Maximum likelihood hyperparameter optimization with respect to the training data was \tcr{used} to optimize any kernel hyperparameters.}
    \label{fig:16_prm_local}
\end{figure*}

Figure \ref{fig:16_prm_local} illustrates this process, showing the log prediction error $\log_{10}(1-R^2_{\mathrm{v}})$ for a repeat of the simulations in figure $\ref{fig:gp_full}$ on reduced parameter scales with scale factors $s=10$ (\ref{fig:16_prm_local}a) and $s=100$ (\ref{fig:16_prm_local}b). 
We see that when required to make only local predictions, GP models using the unitary and classical kernels show significant improvement although those using the state kernel are consistently the most accurate. \tcr{This is to be expected as observations are much denser in the input space. Denser observations allow the universal classical kernel-based models to more effectively interpolate between observations. They also reduce the portion of the state and unitary kernel's induced feature spaces which the observations must span for $k_s$- and $k_u$-based GP models to perform well.}

The state kernel's advantage is most pronounced for $s=10$, which is to be expected as this is closer to global regression. The unitary, RQ, RBF kernels and the Matern kernel with $\nu=5/2$ all perform similarly well but the Matern kernel with $\nu=3/2$ struggles when presented with large amounts of data (particularly for $s=10$). This is likely because a GP equipped with a Matern kernel is $\ceil{\nu}-1$ times differentiable (once differentiable for $\nu=3/2$ and twice for $\nu=5/2$) \cite{williams2006}, while those using the unitary, RBF, and RQ kernels are infinitely differentiable. As the noiseless energy function (a finite weighted sum of Fourier components, see \ref{sec:vqe}) is smooth and infinitely differentiable this may explain why the singly differentiable $\nu=3/2$ Matern kernel performs worse than the other, more-times differentiable kernels.


\section{Bayesian VQE with quantum kernels}
\subsection{Limitations of on-device quantum kernels}
For it to be useful, the predictions from a surrogate model should be significantly easier to compute and optimize than the cost function. Given $m$ observations, making a prediction from a Gaussian process surrogate (see section \ref{sec:GPs}), as is done when optimizing a GP-based surrogate model in BO, costs $m$ kernel evaluations, while fitting the model requires a Gram-matrix of $\mathcal{O}(m^2)$ kernel evaluations to be calculated. 

This raises a potential issue with quantum kernel-based surrogate models, namely that both fitting the surrogate model and using it to make predictions requires repeated \tcr{executions} of a large number of quantum circuits. Unless the number of data points is smaller than the number of circuits required to measure the cost/energy function one could simply measure and optimize the cost directly, avoiding any inaccuracies of the approximate surrogate model.
Accordingly, quantum kernels evaluated on-device are unlikely to be useful for producing the kinds of easy-to-optimize surrogate models needed for Bayesian optimization. 

\tcr{It is also worth noting that unless the ansatz used is highly structured, both the state and unitary kernels evaluated for two randomly chosen sets of gate angles will shrink exponentially with the number of qubits used \cite{fidexpdecay}. While this is less of an issue for small NISQ-scale VQE problems, this means that for problems on many qubits one would require exponentially-many shots to accurately resolve these exponentially-small kernel evaluations. This is closely related to the barren-plateau problem and further reinforces our assertion that quantum kernels evaluated on-device are unlikely to be useful for producing globally-accurate surrogate models for VQE. }

Device noise also greatly complicates the evaluation of quantum kernels on-device. Although using a kernel based on the noisy operation of a quantum processor may be useful in quantifying the similarity between noisy energy evaluations such kernel functions may require large numbers of samples to evaluate accurately. This would greatly complicate any surrogate model optimization and means that the usual requirements of a kernel, for example positivity, would not necessarily be satisfied. It would also be difficult to ensure that the errors present when evaluating the kernel correspond to those encountered when estimating the cost/energy. For example, one typical implementation of the state kernel requires a circuits of twice the depth used to estimate the energy \cite{liu2021} and so involves more noise and decoherence than would be seen in the cost function. Alternatively, one can implement the kernel using a circuit of the same depth as the cost function but over two sets of qubits \cite{cincio2018} which will likely have different noise characteristics.

While Gaussian processes naturally accommodate noisy observations it is normally assumed that the kernel function can be computed exactly. Circuit noise and finite sampling errors would mean an on-device quantum kernel evaluation is affected by statistical fluctuations. This would then affect the two key objects in GP modelling; the positive-definite Gram matrix $\mathbf{K}$ of pair-wise kernel evaluations for the observed points and the vector $\bs{k}$ of kernel evaluations between a point of interest and the observed points (see section \ref{sec:GPs} for a full discussion of these). Randomness in these objects can be detrimental in two ways. 
Firstly, if the Gram matrix $\mathbf{K}$ is composed of noisy kernel evaluations then it may not be positive definite and numerical instabilities can occur when calculating its inverse (needed to make predictions). This can be partially alleviated by increasing the noise strength hyperparameter $\sigma_n$ (see section \ref{sec:GPs}) to ensure an invertible and positive-definite $\mathbf{K}$. 

Another more serious issue comes when attempting to minimize a surrogate model built using a noisy kernel as part of a Bayesian optimization loop. Gradient descent methods are the standard approach for this however they require repeated accurate evaluations of $\nabla_{\bs{x}} \bs{k}$, the gradient of kernel evaluations between the query point and the training data. While noise-resilient methods such as SPSA noise-resilient could be used for this optimization, these could be applied directly to the objective function, bypassing the need for a surrogate model. A possible exception would be if the surrogate model is considerably cheaper to evaluate than the cost function, which would be true if the number of observations is smaller than the number of circuits required to evaluate the cost. 

Due to these complications, we do not expect significant practical advantages to VQE from Gaussian processes which use device-evaluated quantum kernels. Instead we evaluate all the quantum kernels using a classical computer, which can be done tractably provided the number of qubits is relatively small and depth of the quantum circuits involved is relatively low.

\tcr{For many-qubit VQE problems, in which most kernel evaluations are exponentially suppressed, one could still attempt a highly localized form of Bayesian VQE with classically-evaluated quantum kernels. In such scenarios, the Gram matrix $\mathbf{K}$ calculated for a set of randomly chosen initial points will be exponentially close to the identity. Similarly, kernel evaluations between a new point of interest $\ptheta^*$ and these initially chosen points will be close to zero unless $\ptheta^*$ is in the immediate vicinity of an initial point. Bayesian VQE using a GP with such an exponentially-decaying kernel will be highly localised around whichever initial point gives the lowest energy. This means that at each step in the BO, the maximum of the acquisition function (see section \ref{sec:BO}) will stay close to the most promising initial point (where the kernel is non-negligible). This optimization strategy would still differ from completely local methods like as gradient descent in that all previous observations (however close to each other they may be) are used to decide the next query point rather than just the most recent point seen. However for this to work, one would still need to find a useful starting point for the optimization (i.e. not in a barren-plateau) and have a method for evaluating the kernel. This could either be estimated on-device, would likely be more costly than optimizing the energy directly, or evaluated classically, which would be intractable unless the ansatz has some simple exploitable structure or admits a simplifying approximation (we discuss this in section \ref{sec:approx}).}


\subsection{Noiseless Bayesian VQE}\label{sec:bayesian_vqe}
We have seen that the state kernel provides a significant advantage over typical classical kernels when performing GP regression of a circuit's energy landscape and now apply these results to VQE using Bayesian optimization. Bayesian optimization is a gradient-free strategy for optimizing noisy expensive-to-evaluate objective functions \cite{snoek2012}. As current cloud-based NISQ computers are in high demand, the problem of variationally minimizing the noisy $\tilde{E}(\ptheta)$ is an apt use-case for Bayesian optimization. We describe Bayesian optimization in more detail in section \ref{sec:BO}. 

\begin{figure*}
    \centering
    \includegraphics[width=\linewidth]{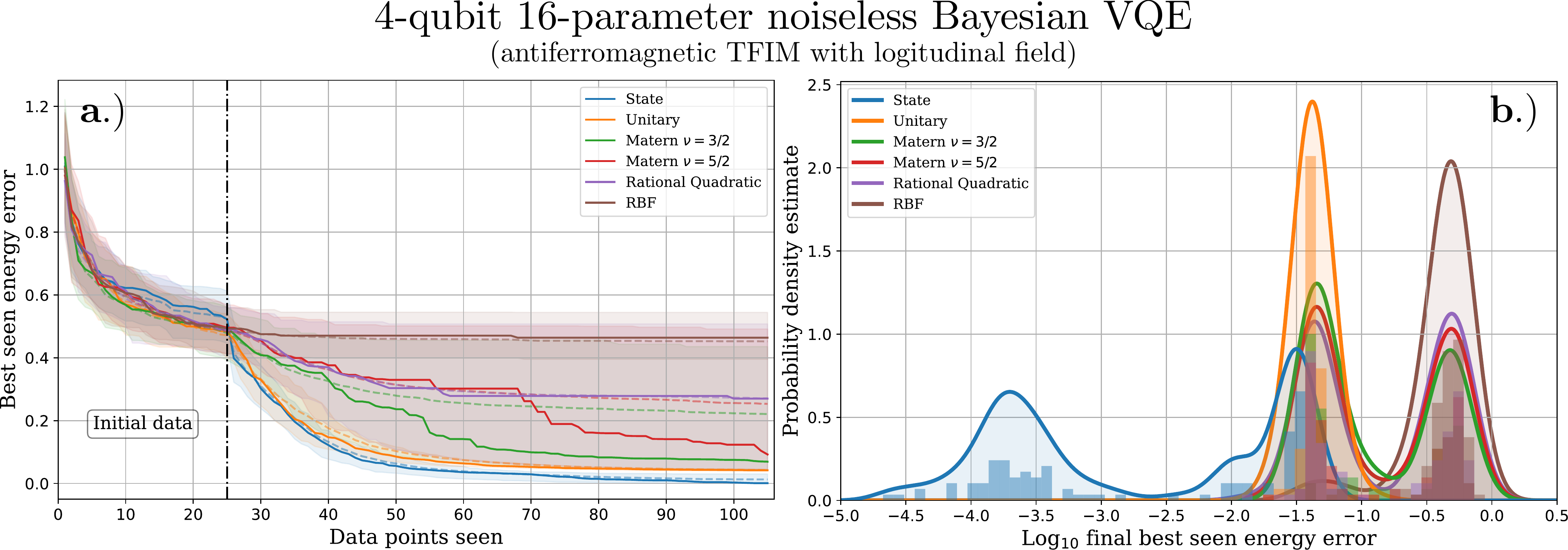}
    \caption{\textbf{Performance of 16 parameter noiseless Bayesian VQE using different kernels.} The Hamiltonian is given in \eqref{eq:ham} with $J=h_z=0.5$ and $h_x=-0.5$, while the ansatz used here is the 16-parameter 4-qubit ansatz in Figure \ref{fig:ansatz}. We use the Expected Improvement acquisition function with $\xi=0.01$ and maximum likelihood hyperparameter optimization. All kernels were equipped with a signal variance hyperparameter $\sigma^2$, i.e. $k(\bs{\theta},\bs{\theta}')\to\sigma^2k(\bs{\theta},\bs{\theta}')$.
    \textbf{a.)} Vertical axis is the best seen energy error \eqref{eq:energy_error}, horizontal axis is the number of data points seen. The median (solid lines), mean (dashed lines), and interquartile ranges (filled) are shown for data aggregated over 100 repeats of the optimization. The first 25 points are chosen at random to initialize the optimization (indicated by black vertical dashed line).
    \textbf{b.)} Horizontal axis is the log final best seen energy error $\log_{10}\mathcal{E}(\bs{y},\ptheta_{\mathrm{opt}})$ after 80 points have been queried. Vertical axis is a Gaussian kernel density estimate (bandwidth 0.15) of the distribution of $\log_{10}\mathcal{E}(\bs{y},\ptheta_{\mathrm{opt}})$ (solid line and filled), estimated from the 100 repeats. Also shown are histograms of the final best seen energy error for the 100 repeats (solid bars).}
    \label{fig:noiseless_vqe}
\end{figure*}
We will quantify the performance of the optimization in terms of the best seen energy error $\mathcal{E}(\bs{y},\ptheta_{\mathrm{opt}})$ which we define as the fractional difference between the lowest energy seen at the current stage in the optimization and the minimum achievable noiseless energy for the ansatz. For a set of observed energy values $\bs{y}$ and an optimal set of gate angles $\ptheta_{\mathrm{opt}}$, as defined in \eqref{eq:vqe_goal}, this is given by
\begin{equation}\label{eq:energy_error}
    \mathcal{E}(\bs{y},\ptheta_{\mathrm{opt}}) \coloneqq \frac{\min(\bs{y}) - E(\ptheta_\mathrm{opt})}{\abs{E(\ptheta_\mathrm{opt})}}.
\end{equation}
The optimal gate angles $\ptheta_{\mathrm{opt}}$ for the ansatz shown in Figure \ref{fig:ansatz} were found using $10,000$ attempts of direct gradient-based minimization, giving a minimum energy of $E(\ptheta_{\mathrm{opt}})=-2.762194$. 

Figure \ref{fig:noiseless_vqe} shows the results of noiseless VQE simulations using Bayesian optimization with different kernels. The ansatz circuit used is shown in Fig \ref{fig:ansatz} and the Hamiltonian is given in \eqref{eq:ham} with $J=h_z=0.5$ and $h_x=-0.5$. At the start of each simulation, $25$ initial points were drawn uniformly at random and their energies evaluated to form an initial training data set. The Expected Improvement \cite{jones1998} acquisition function was used with an exploration hyperparameter $\xi=0.01$, this value is widely used in EI-based BO \citep{Lizotte2008,Brochu2010} and was found to yield the best BO performance across the classical kernels considered. For each kernel, 100 repeats of the optimization were performed. Figure \ref{fig:noiseless_vqe}a shows the median (solid lines), mean (dashed lines), and inter-quartile (filled) range of the energy error. Figure \ref{fig:noiseless_vqe}b gives histograms and Gaussian kernel density estimates (with bandwidth $0.15$ \cite{scikit-learn}) of the final best seen energy errors (after 80 new points have been requested). 

The state kernel outperforms the other kernels at this VQE task, both in terms of the final energy found and in the speed at which it converges. It frequently reaches an energy less than one part in $10^{-3.5}\approx 0.03\%$ away from the minimum achievable with this ansatz. Its final achieved energies form clusters around a few values, the most noticeable being one at errors between $10^{-4}$ and $10^{-3.5}$ and one that is close to, but still lower than, the majority of energies observed with the unitary and classical kernels. 
The unitary kernel also performs well, quickly and consistently converging to an error of approximately $4\%$ and exhibits the second lowest mean and median errors overall. However, the data is strongly clustered with this error, implying that achieving a more accurate final energy may be difficult. The classical kernels vary considerably in their performance, generally having much worse mean errors than the quantum kernels. The final energies achieved by the classical kernels form two clusters; one of relatively successful runs with errors $\sim 6\%$ and one with errors between $30-60\%$. The errors in this latter cluster are similar to those seen in the randomly selected initialization data, implying that these optimization runs failed to achieve any significant reduction in the energy. 
This suggests that while these GP models using these classical kernels sometimes lead to relatively successful VQE, they often immediately and become stuck in an exploitative phase where points close to those in the initial training data are repeatedly queried.
It is therefore likely that as well as their advantages in over-all performance, Bayesian VQE using the state or unitary kernel can be more resilient against initialization failures than when using the classical kernels we have considered. 

\begin{figure*}
    \centering
    \includegraphics[width=\linewidth]{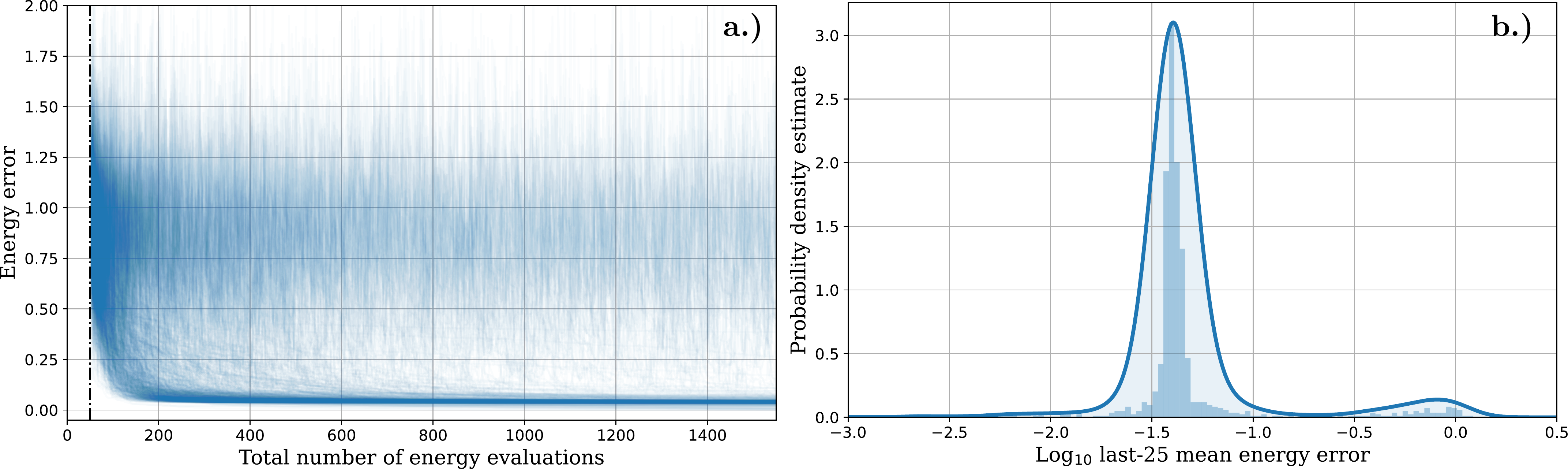}
    \caption{\textbf{Performance of 16 parameter noiseless VQE using SPSA.} The Hamiltonian is given in \eqref{eq:ham} with $J=h_z=0.5$ and $h_x=-0.5$. The ansatz used is the 16-parameter 4-qubit ansatz in Figure \ref{fig:ansatz}. We used the Qiskit SPSA implementation \citep{qiskit,spall1992} in which 50 initial energy evaluations are used to calibrate the optimizer. The initial ansatz parameters for each run were chosen uniformly at random.
    \textbf{a.)} Vertical axis is the energy error \eqref{eq:energy_error}, horizontal axis is the number of energy evaluations at current point in optimization (including those to estimate the gradient). Data for 1,000 repeats are shown. 
    \textbf{b.)} Horizontal axis is the logarithm of the mean energy error for the last $25$ energy evaluations in each optimization run. Vertical axis is a Gaussian kernel density estimate (bandwidth 0.1) of the distribution of this data (solid line and filled), estimated from the 1,000 repeats. Also shown are histograms of the data for the 1,000 repeats (solid bars).}
    \label{fig:SPSA}
\end{figure*}

For comparison, Figure \ref{fig:SPSA} shows a repeat of this noiseless VQE using the SPSA optimization scheme. We used the Qiskit implementation \cite{qiskit} which closely follows the initial proposal in \cite{spall1992}. We see from the experimental traces \ref{fig:SPSA}a and histograms in \ref{fig:SPSA}b that the effectiveness of this optimization strategy can vary but the average final energies typically have errors in line with those obtained for BO using the unitary kernel at around $4\%$. These are also comparable to the best experimental runs for BO using the various classical kernels. The data is strongly clustered around this energy error and only a handful of SPSA runs manage to compete with the low energies frequently achieved with state kernel-based BO. The SPSA runs which cluster around $4\%$ generally require several hundred energy evaluations to achieve this level of accuracy whereas only around $100$ evaluations are needed for the various kernel based models. The two SPSA runs with the lowest mean final energies had mean final energy errors on the order of $0.1\%$ which is comparable to the lowest energies achieved with state kernel-based BO. However, these were only seen after $1,500$ total energy evaluations to be made. This level of final accuracy with SPSA appears to be extremely rare and requires greater than an order of magnitude more energy evaluations than is needed with the state kernel, highlighting the effectiveness and economy of our optimization strategy.

\subsection{Noisy Bayesian VQE}

The question remains whether a GP surrogate using a noiseless classically-evaluated quantum kernel is useful for Bayesian VQE of noisy quantum circuits. The noisy energy function is unlikely to be a linear function in the feature spaces induced by the noiseless state and unitary kernels. However, provided the device noise is not too great, these noiseless classical evaluations should provide a good approximation to the correct feature space for describing the noisy energy. 
By using a quantum kernel function based on the ansatz circuit we are able to leverage our prior knowledge of the (noiseless) ansatz circuit whereas standard GP surrogates can only attempt to do this through hyperparameter optimization. 
Because the observed energy values provide us with information about the noise on the device we are also able to take this noise into account implicitly. 

\begin{figure*}[!ht]
    \centering
    \includegraphics[width=\linewidth]{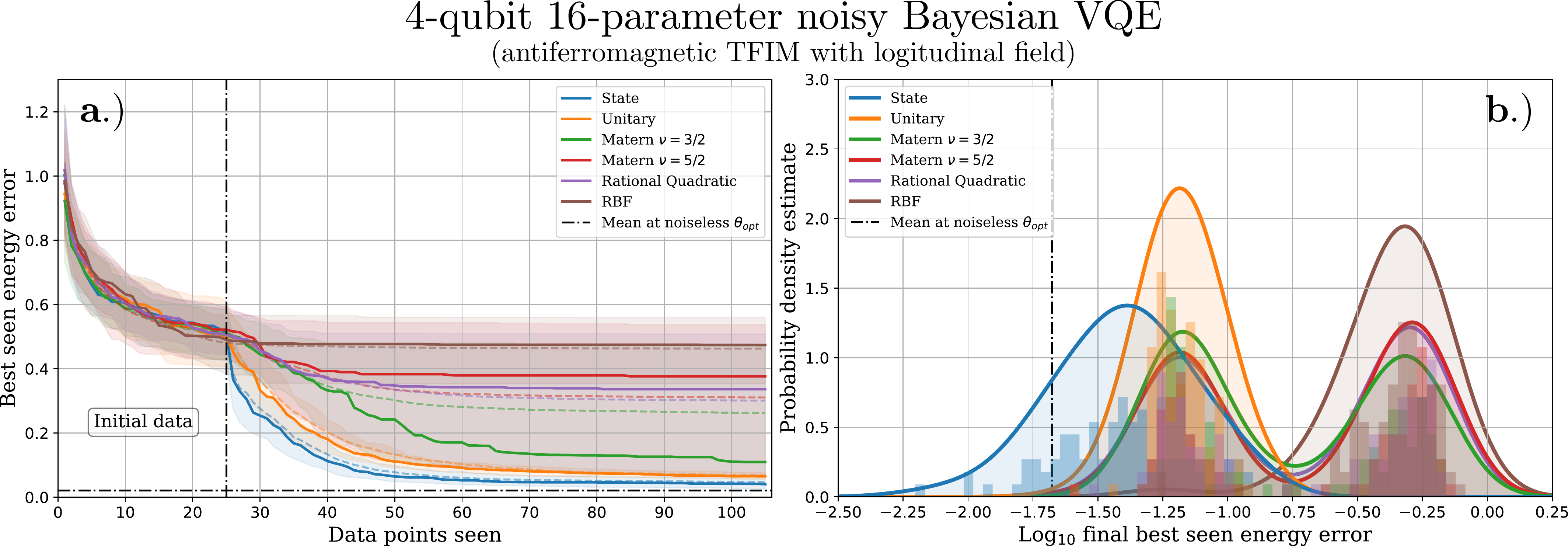}
    \caption{\textbf{Performance of 16 parameter noisy Bayesian VQE using different kernels.} The Hamiltonian is given in \eqref{eq:ham} with $J=h_z=0.5$ and $h_x=-0.5$, while the ansatz used here is the 16-parameter 4-qubit ansatz in Figure \ref{fig:ansatz}. The noise model used was derived from the errors present on $\textit{ibmq\_quito}$ in October 2021 \cite{Quito}. We use the Expected Improvement acquisition function with $\xi=0.01$ and maximum likelihood hyperparameter optimization. All kernels were equipped with a signal variance hyperparameter $\sigma^2$ and noise hyperparameter $\sigma^2_n$, i.e. $k(\bs{\theta}_i,\bs{\theta}_j)\to\sigma^2k(\bs{\theta}_i,\bs{\theta}_j) + \sigma_n^2\delta_{ij}$
    .
    \textbf{a.)} Vertical axis is the best seen energy error \eqref{eq:energy_error}, horizontal axis is the number of data points seen. The median (solid lines), mean (dashed lines), and interquartile ranges (filled) are shown for data aggregated over 100 repeats of the optimization. The first 25 points are chosen at random to initialize the optimization (indicated by black vertical dashed line). 
    \textbf{b.)} Horizontal axis is the log final best seen energy error $\log_{10}\mathcal{E}(\bs{y},\ptheta_{\mathrm{opt}})$ after 80 points have been queried. Vertical axis is a Gaussian kernel density estimate (bandwidth 0.15) of the distribution of $\log_{10}\mathcal{E}(\bs{y},\ptheta_{\mathrm{opt}})$ (solid line and filled), estimated from the 100 repeats. Also shown are histograms of the final best seen energy error for the 100 repeats (solid bars).
    The black horizontal dashed line in \textbf{a} and vertical dashed line in \textbf{b} shows the mean of $10,000$ evaluations of $\tilde{E}(\bs{\theta}_{\mathrm{opt}})$, the noisy energy at the noiseless optimal ansatz parameters.}
    \label{fig:noisy_vqe}
\end{figure*}

Figure \ref{fig:noisy_vqe} shows the results of a repeat of the Bayesian VQE simulations illustrated in Figure \ref{fig:noiseless_vqe} with noisy energy evaluations. These were performed with using Qiskit's noisy quantum circuit simulation framework \cite{qiskit} and a noise model derived from the errors present on $\textit{ibmq\_quito}$ in Oct 2021 \cite{Quito}. The noise includes contributions from gate errors, state preparation and measurement errors (for which Qiskit's readout error mitigation was used), and finite circuit shots (10,000 per circuit per evaluation). The quantum kernel evaluations (where used) were simulated exactly using a tensor network quantum circuit implementation based on the Quimb Python package \cite{gray2018}. As in Figure \ref{fig:noiseless_vqe}, Figure \ref{fig:noisy_vqe}a shows the median (solid lines), mean (dashed lines), and inter-quartile range (filled) of the best seen energy error; the difference between the lowest seen energy and the minimum possible for the ansatz. Figure \ref{fig:noisy_vqe}b shows histograms and kernel density estimates (with bandwidth 0.15) of the final best seen energy errors (after 80 new points are queried). 

The noise in the simulations makes it unlikely that the energy can reach the noiseless minimum value. Accordingly, we also show the mean energy error obtained from $10,000$ repeated noisy evaluations of the energy $\tilde{E}(\ptheta_{\mathrm{opt}})$, taken at $\ptheta_{\mathrm{opt}}$, the noiseless optimum gate angles. While $\ptheta_{\mathrm{opt}}$ may differ slightly from the gate angles that yield the true noisy optimum these evaluations serve as an indicator of the best performance to be expected from a noisy VQE implementation.

We again see the Gaussian process models based on the state kernel greatly outperform those using the other kernels. The state kernel frequently manages to achieve energies lower than the mean $\tilde{E}(\bs{\theta}_{\mathrm{opt}})$ (an error of $\sim2\%$) while also showing the fastest convergence. The unitary kernel also performs well, consistently giving energies lower or at least as low as those found using classical kernels (around $6\%$) but with a significantly lower variation in its performance.
When dealing with noisy evaluations, VQE using the classical kernels shows qualitatively similar performance to the noiseless case (although with different final energy errors). Almost all the simulations performed with the RBF kernel and approximately half of those with the other classical kernels give a final errors in the range $30-60\%$. As in the noiseless case, these errors are similar to those seen for the randomly-selected initialization points implying that these runs failed almost immediately. The remainder of the classical kernel simulations achieve a much better performance which is close to the $\sim 6\%$ seen with the unitary kernel.

\section{Classical simulation of quantum kernels}\label{sec:approx}
We have demonstrated that Bayesian VQE is significantly more effective when using quantum kernel-based GP models than when using classical kernels, both in terms of final energy accuracy and reliability. However, our simulations have concerned circuits of sufficiently few qubits and low depth that classical simulation of the quantum kernels is tractable. Once the number of qubits and/or the depth of the variational ansatz becomes too large, a GP surrogate model built on the full quantum kernel would be too computationally expensive to be practically used. However, if the number of gates to be optimized is not too great, in some instances classical evaluation of quantum kernels can remain tractable even for large numbers of qubits. 

One way to ensure this is to only perform optimization on a subset of gates at any given time. Suppose we have an ansatz of the form $U(\ptheta)=U_C(\ptheta_C)U_B(\ptheta_B)U_A(\ptheta_A)$, where $U_C$, $U_B$, and $U_A$ do not commute and any constant gates which commute with $U_B$ have been included in $U_A$ or $U_C$. If, at some stage in the VQE, we fix $\ptheta_A$ and $\ptheta_C$ and only optimize over $\ptheta_B$ then the unitary kernel is given by $k_u(\ptheta_B,\ptheta_B')=\abs{\Tr{U^\dagger_B(\ptheta_B')U(\ptheta_B)}/d}^2$. Both the gates in the past ($U_A$) and future ($U_C$) causal light-cones of $U_B$ cancel out due to the cyclic property of the trace meaning that the classical computational cost of evaluating $k_u(\ptheta_B,\ptheta_B')$ only depends on the complexity of $U_B$. 

However, as we have seen in the previous sections, while the unitary kernel does provide some advantages over the typical classical kernels we have considered, it usually requires many more observed energy evaluations than the state kernel to produce a similarly accurate surrogate model. The state kernel for this block-wise parametrization is given by $k_s(\ptheta_B,\ptheta_B')=\abs{\bra{0}U_A^\dagger(\ptheta_A)U^\dagger_B(\ptheta_B')U(\ptheta_B)U_A(\ptheta_A)\ket{0}}^2$. For this kernel, the gates in the past causal light-cone of $U_B$ (those in $U_A$) must still be retained. Equivalently the kernel can be written as
\begin{equation}\label{eq:limited_skern}
    k_s(\ptheta_B,\ptheta_B')=\abs{\bra{\psi_A}U^\dagger_B(\ptheta_B')U(\ptheta_B)\ket{\psi_A}}^2,
\end{equation}
where $\ket{\psi_A}=U_A(\ptheta_A)\ket{0}$ is the input state for an equivalent state kernel only containing $U_B$. If $U_A$ is a deep circuit across many qubits such that $\ket{\psi_A}$ cannot be represented or manipulated in a classically efficient manner, then this could make classical evaluation impractical and so prevent the kernel's use in a classical GP model.

Block or layer-wise VQE as we describe above has received significant attention in the literature as a potential solution to the barren-plateau issues that prevent large-scale VQE \citep{parrish2019,Slattery2021,Skolik2021}. If we assume that VQE can be performed effectively by selecting small blocks or layers of gates ($U_B$) to optimize at a time then one could attempt to find a classically tractable approximation to $\ket{\psi_A}$ in order to produce an approximate state kernel. In section \ref{sec:mps} we describe a scheme that approximates $\ket{\psi_A}$ to a matrix product state (MPS).

Matrix product states are a classically-efficient representation for certain types of multi-partite quantum state \citep{Garcia2007,Verstraete2008}. They first saw widespread use in condensed matter physics as the underlying ansatz for the powerful density matrix renormalization group (DMRG) algorithm (and other closely related algorithms) for 1-dimensional quantum lattices \cite{Schollwock2011} and are an example of the wider class of tensor network states \cite{Orus2014}. 
A matrix product state (with periodic boundary conditions) on $n$ sites is defined in terms of a collection of $n$ rank-3 tensors with a single outgoing physical index (of dimension equal to that of the subsystem at the site) and one or two (depending on periodicity) virtual indices. The state is given as a contraction over pairs of virtual indices for adjacent sites, forming bonds between them and leaving only the physical indices.
The amount entanglement present in the state and the complexity of representing and manipulating the MPS depends on the size of the bond indices.

Our scheme approximates the input state $\ket{\psi_A}$ by starting with an MPS representation of the input state $\ket{\bs{0}}$ (starting with all bond dimensions equal to $1$) and contracting each constant gate from $U_A$ into this MPS. Single qubit gates can be contracted into an MPS without changing its bond dimensions. To apply a two-qubit gate to neighbouring sites/qubits in an MPS one first contracts the gate unitary into the tensors at the two involved sites. This yields a new two-site tensor which is then broken back down into two single-site tensors using singular-value-decomposition (SVD). 
These new MPS tensors are connected by a bond with dimension at most a factor of $4$ larger than the original bond between the sites. 
To ensure the bond dimensions do not grow exponentially as more gates are applied, the size of the final new bond is capped at a fixed $\chi_{\mathrm{max}}$ by retaining only the largest $\chi_{\mathrm{max}}$ singular values of the two-site tensor (and the resulting state is re-normalized). Discarding singular values in this way is likely to reduce the fidelity between the truncated state and the state without truncation, degrading the accuracy of the overall approximation. To mitigate this decay in accuracy the truncated tensors are optimized to maximise the fidelity between the state with and without truncation. After all gates have been applied in this way we arrive at a new input state $\ket{\tilde{\psi}_A^{(\chi_\mathrm{max})}}$ which is an MPS with a maximal bond dimension $\chi_\mathrm{max}$ which we then use to obtain an approximation to the $k_s$ given by
\begin{widetext}
\begin{equation}
   \tilde{k}_s(\ptheta_B,\ptheta_B',\chi_\mathrm{max}) \coloneqq \abs{\bra{\tilde{\psi}_A^{(\chi_\mathrm{max})}}U^\dagger(\ptheta_B')U(\ptheta_B)\ket{\tilde{\psi}_A^{(\chi_\mathrm{max})}}}^2.
\end{equation} 
\end{widetext}

If $U_B$ is only finitely entangling (being either low width or low depth) it can be contracted into $\ket{\tilde{\psi_A}}$ with only a finite multiplicative increase in the MPS's bond-dimension. This means that if $\chi_{\mathrm{max}}$ is kept relatively small we can evaluate the approximate state kernel at a cost that scales at most as $\mathcal{O}(n\chi_{\mathrm{\max}}^3 d)$, where $d=2$ is the individual physical dimension of the $n$ qubits \cite{Paeckel2019}.

While our proposed scheme approximates the input state to a MPS \tcr{this is by no means the only classically-efficient representation one could use to approximate} $\ket{\psi_A}$. For simplicity we only consider nearest-neighbour entangling gates; two-qubit gates between non-neighbouring qubits can also be contracted into an MPS state by means of additional SWAP gates at the cost of large increases in the bond dimensions. Depending on the situation and structure of $U_A$, one could instead attempt to use one of the numerous other classically-efficient tensor network states. For example, a more complicated 2- (or higher-) dimensional entangling structure may motivate the use of projected entangled pair states (PEPS) or tree tensor-networks \cite{Bridgeman2017}.

By approximating $\ket{\psi_A}$ gate-by-gate with an upper limit on the retained bond-dimensions, our scheme avoids explicitly calculating the potentially intractable $\ket{\psi_A}$. However, truncating the bond-dimension inevitably leads to a reduction in the fidelity between $\ket{\psi_A}$ and $\ket{\tilde{\psi}^{(\chi_\mathrm{max})}_A}$. Instead, one could choose a classically-efficient parameterization of the input state $\ket{\psi_A}$ as a kernel hyperparameter to optimize. This would allow an input state to be chosen with reference to the observed data, rather than being the result of a series of approximations. However, optimizing only a handful of kernel hyperparameters can be extremely expensive. As we discuss in section \ref{sec:GPs}, optimizing the marginal likelihood with respect to any hyperparameter involves repeated calculation of both the Gram matrix $\mathbf{K}$, its inverse, and its derivative (see \eqref{eq:gradml}). Unless the representation of $\ket{\psi_A}$ is extremely (hyper)parameter efficient, this would likely make such a scheme impractical. 

\begin{figure*}
    \centering
    \includegraphics[width=\linewidth]{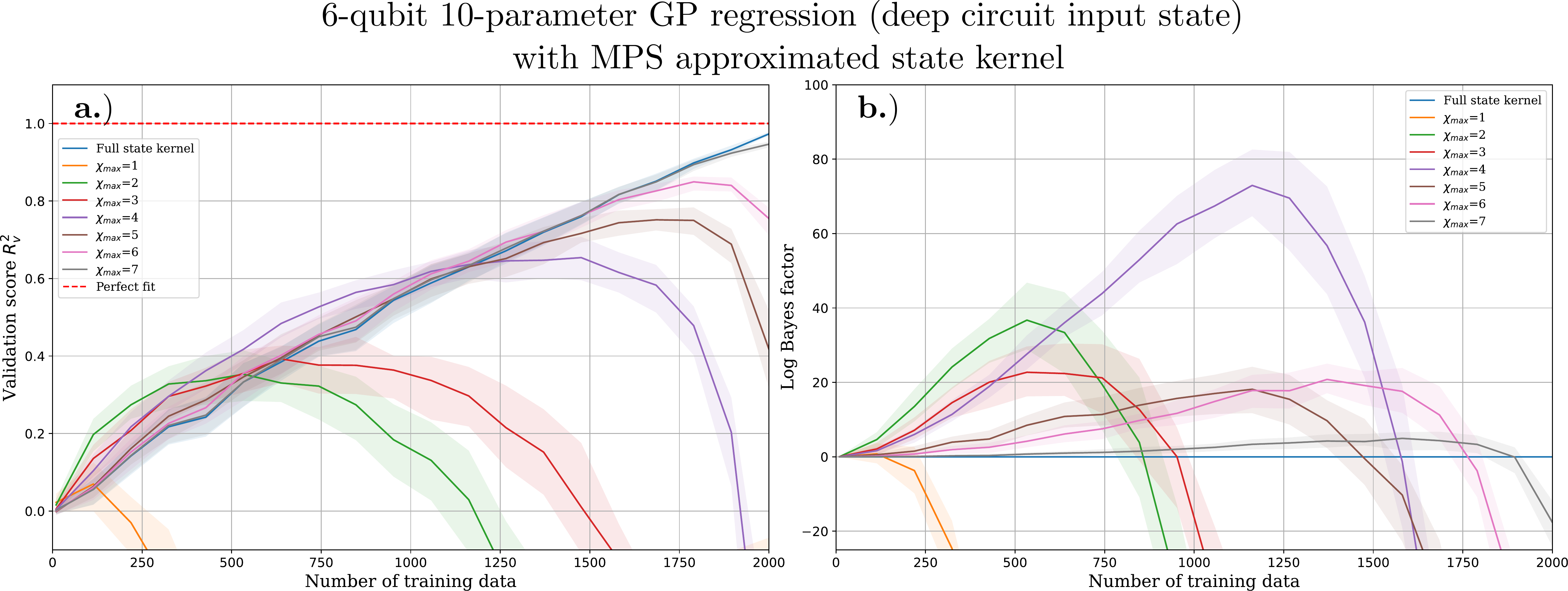}
    \caption{\textbf{Validation scores $R^2_{\mathrm{v}}$ and log Bayes factors $\log\mathcal{B}$ for GP regression models with approximated state kernels of various bond dimensions learning the energy landscape $E(\bs{\theta})$ for the Hamiltonian in \eqref{eq:ham} with $J=h_z=0.5$ and $h_x=-0.5$.} The ansatz had the same structure as that in Figure \ref{fig:ansatz} but involved six qubits and a circuit depth of 20. Only the last two layers of $RY$ gates were varied (a $U_B$ with 10 parameters in total) and the remainder of the circuit ($U_A$) was initialized with uniform random gate angles and held constant. A classically-efficient approximation to the state kernel was produced by approximating the state $\ket{\psi_A}=U_A\ket{0}$ before the parameterized $U_B$ to a matrix product state with maximal bond dimension $\chi_\mathrm{max}$. Results are also shown for simulations using the full un-approximated state kernel. The training and validation points are chosen uniformly at random (for each of the 10 parameters) in the range $[-\pi,\pi]$ and the corresponding energies $\bs{y}_{\mathrm{t}}$ and $\bs{y}_{\mathrm{v}}$ are evaluated noiselessly on a classical computer. \textbf{a.)} Horizontal axis is the size of the training data set $N_{\mathrm{t}}$, vertical axis is the validation score (defined in \eqref{eq:val_score}). \textbf{b.)}  Horizontal axis is the size of the training data set $N_{\mathrm{t}}$, vertical axis is the log Bayes factor (defined in \eqref{eq:lbf}) versus the full un-approximated state kernel for the different approximated kernels.
    In both plots we show the median over 100 repeats (solid lines/dots) and the inter-quartile range of the data (filled).}
    \label{fig:mps_plot}
\end{figure*}

\subsection{Gaussian process regression with an MPS-approximated state kernel}
To test our approximation strategy we compare the predictive accuracy of a GP model built with approximated state kernels of various $\chi_{\mathrm{max}}$ to the accuracy when using the full state kernel. We again \tcr{consider} the Hamiltonian in \eqref{eq:ham} with $J=h_z=0.5$ and $h_x=-0.5$. To ensure that the full state kernel is tractable for large numbers of observations we use a brickwork ansatz with the same structure and open boundary conditions as in \eqref{fig:ansatz} but with 6 qubits \tcr{and} 20 layers of $CX$ gates on alternating pairs of adjacent qubits separated by $RY$ gates (a total of 50 $CX$ gates and 106 $RY$s). To engineer a situation in which a low $\chi_{\mathrm{max}}$ MPS approximation is unlikely accurate we choose the parameterized portion of the ansatz, $U_B$, as the last two layers of $RY$s (10 of which surround the final $CX$ layer) and pick a fixed set of random gate angles for the remaining gates (which form $U_A$). For each simulation we generate two sets of random uniform gate angles $\mathbf{X}_{\mathrm{t}}$ and $\mathbf{X}_{\mathrm{v}}$ and their corresponding noiseless energies $\bs{y}_{\mathrm{t}}$ and $\bs{y}_{\mathrm{v}}$ as training and validation data. The size of the training data set is varied while the size of the validation set is fixed to $N_{\mathrm{v}}=100$.

Figure \ref{fig:mps_plot}a shows the validation scores achieved by Gaussian process surrogates equipped with MPS approximated state kernels of different $\chi_{\mathrm{max}}$ compared to one equipped with the full state kernel. The fidelities of the MPS approximations with different maximum bond dimensions to $\ket{\psi_A}$ are shown in table \ref{tab:mps_fid}. To help assess the relative suitability of the different GP models in explaining their training data, Figure \ref{fig:mps_plot}b shows the Log Bayes factor between the GP models with different approximated kernels and a GP model using the full state kernel. The Bayes factor, $\mathcal{B}(\mathbf{X},\bs{y},A,B) = { p(\bs{y}|\mathbf{X},A)}/{p(\bs{y}|\mathbf{X},B)}$, \cite{Kass1995} between two statistical models $A$ and $B$ is defined as the ratio of the marginal likelihoods of a given set of training data under the two models. Its logarithm is
\begin{equation}\label{eq:lbf}
    \log \mathcal{B}(\mathbf{X},\bs{y},A,B) = M_A(\mathbf{X},\bs{y}) - M_B(\mathbf{X},\bs{y}),
\end{equation}
where $M_A(\mathbf{X},\bs{y})$ and $M_B(\mathbf{X},\bs{y})$ are the log marginal likelihoods of the data $(\mathbf{X},\bs{y})$ with the models $A$ and $B$ respectively (which for the GP models we use is given by \eqref{eq:ml}). 
If the models $A$ and $B$ are assigned equal prior probabilities then the Bayes factor is equivalent to the ratio of the posterior probabilities of the two models $p(A|\bs{y},\mathbf{X})/p(B|\bs{y},\mathbf{X})$. This gives the Bayes factor a useful interpretation as how much more plausible one model is at explaining the data than the other \cite{Kass1995}. A log Bayes factor $\mathcal{B}(\mathbf{X},\bs{y},A,B)>1$ generally implies that model $A$ is more strongly supported by the data than model $B$ \cite{Kass1995}.

We see that MPS approximations to $\ket{\psi_A}$ with a higher $\chi_{\mathrm{max}}$ have a larger fidelity with $\ket{\psi_A}$ state and their GP models show increasingly similar performance to the GP which uses the full state kernel. Interestingly, the GP models that use approximated state kernels with lower $\chi_{\mathrm{max}}$ often have a higher validation score and Bayes factor (relative to the full state kernel) $\gg1$, provided the size of the training data set is small. When dealing with small numbers of observations these models have both a higher predictive accuracy for unseen energies and are better supported by their training data than the model using $k_s$. This may be because the induced feature spaces of these kernels have a smaller dimension than the full state kernel, being based on a restricted subspace of the $n$ qubit Hilbert space. As a result, less data would be needed to span an appreciable portion of the feature space and so make accurate predictions.

\tcr{For all the approximated kernels there comes a point where the validation score begins to decrease as more training data is presented. 
This happens when the size of the training data set becomes close to the dimension of the approximated kernel's feature space. 
As the energy function will not be completely linear in the approximated kernel's feature space, it becomes increasingly difficult to reconcile these different observations well with a linear model.
This feature space saturation will also occur for the GP based on the full state kernel when the validation score reaches $R^2_{\mathrm{v}}=1$ and the data spans the whole feature space. However, this is not an issue because the energy function is truly linear in the full state kernel's feature space and so energy observations can always be fully reconciled by a linear model in this space. 
Linear dependence will occur when the number of observations exceeds the feature space dimension at which point the Gram matrix $\mathbf{K}$ also becomes rank deficient; even before this point small eigenvalues will be present in $\mathbf{K}$ causing the GP model's predictions to become numerically unstable.}

\tcr{Typically a small regularizing offset, e.g. $\mathbf{K}\to \mathbf{K}+10^{-10}\mathbf{I}$, is added to Gram matrices to ensure they are invertible by effectively adding additional dimensions to the feature vectors. 
Provided the offset is small, the regularizaiton has little effect on models with well-suited kernels, like the full state kernel, as the objective function has minimal dependence on the additional components of the feature vectors introduced by the regularization. 
It can however help smooth out apparent inconsistencies encountered when using a kernel whose feature space is improperly aligned with the objective function by increasing the small eigenvalues present in $\mathbf{K}$ and reducing their disproportionate contribution to the predictions.}

\begin{table*}[]
    \centering
    \begin{tabular}{||c||c|c|c|c|c|c|c|}
    \hline
         $\chi_{\mathrm{max}}$&1&2&3&4&5&6&7  \\
         \hline
        Fidelity & 0.0054& 0.2904& 0.5788& 0.8341& 0.9382& 0.9716& 0.9976 \\
    \hline
    \end{tabular}
    \caption{\textbf{Fidelities of MPS approximations with $\ket{\psi_A}$ used in simulations.} For the maximum bond dimensions allowed in the MPS approximation, the fidelity $\abs{\braket{\psi_A}{\tilde{\psi}_A^{(\chi_{\mathrm{max}})}}}^2$ with the unapproximated $\ket{\psi_A}$ is shown.}
    \label{tab:mps_fid}
\end{table*}

\tcr{Choosing the value for this offset is key to ensuring accurate and numerically stable predictions; it must be sufficiently large to damp the contributions of small eigenvalues of $\mathbf{K}$ but not so large that it dominates the model. We see in Figure \ref{fig:mps_plot}b that when the feature space saturation occurs, degrading the validation score, the likelihood of the training data also drops precipitously. As this can be calculated from the training data alone, we can use this as a metric for calibrating the diagonal offset to avoid numerical instabilities. We do this by treating the weighting of the diagonal offset as a kernel hyperparameter $\sigma_n^2$ such that $\mathbf{K}\to\mathbf{K}+\sigma_n^2 I$ and optimizing this to maximize the marginal likelihood of the observations. 
Adding this hyperparameter is equivalent to assuming that we are dealing with observations corrupted by additive Gaussian noise with zero mean and variance $\sigma_n^2$; discrepancies between the observations and the assumption that they are drawn from a linear function in the kernel's feature space are effectively treated as an additional source of noise. When dealing with real noisy data, adding this noise term to avoid feature space saturation issues introduces negligible additional cost as a this term can be used to simultaneously deal with the observation noise. }

\begin{figure*}
    \centering
    \includegraphics[width=\linewidth]{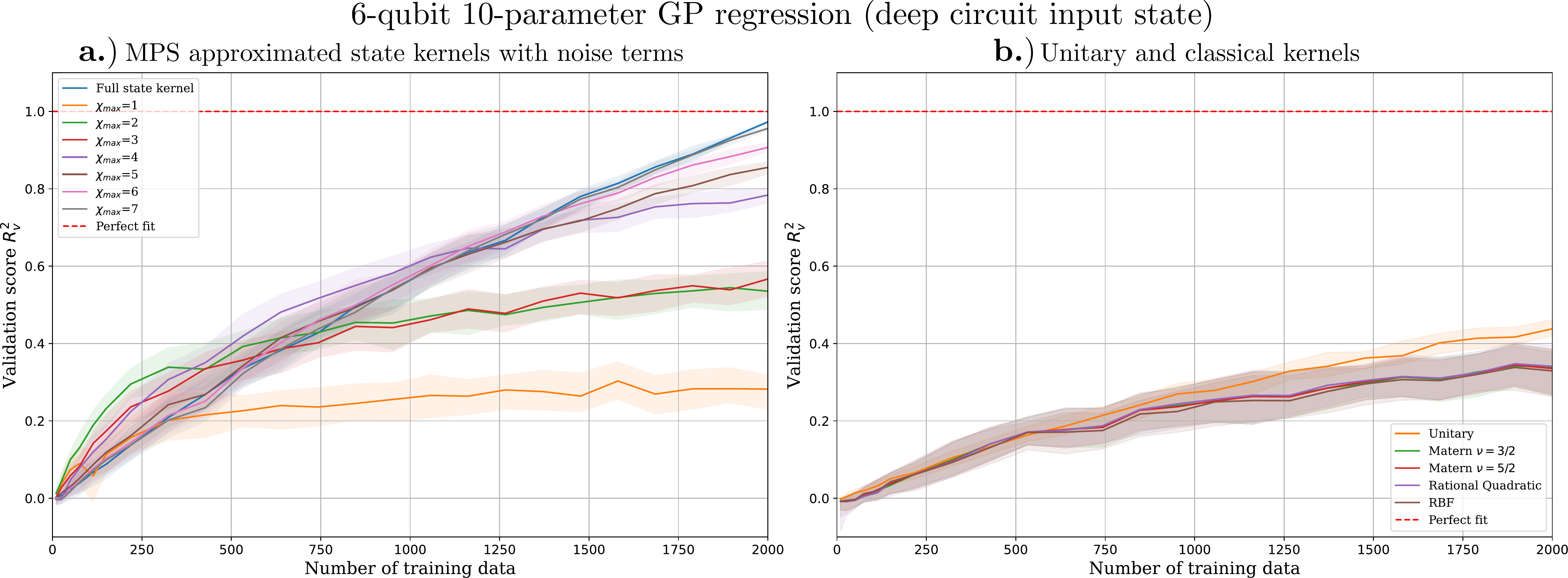}
    \caption{\textbf{Validation scores $R^2_{\mathrm{v}}$ for GP regression models based on approximated (with added noise terms) and full quantum kernels and classical kernels, learning the energy landscape $E(\bs{\theta})$ for the Hamiltonian in \eqref{eq:ham} with $J=h_z=0.5$ and $h_x=-0.5$.} The ansatz and parameterization used was the same as in Figure \ref{fig:mps_plot}. The approximated state kernels were the same as in Figure \ref{fig:mps_plot} but were given an extra noise term $\tilde{k}_s(\ptheta_i,\ptheta_j)\to\tilde{k}_s(\ptheta_i,\ptheta_j)+ \sigma_n^2\delta_{ij}$ to avoid kernel saturation issues. The training and validation points are chosen uniformly at random (for each of the 10 parameters) in the range $[-\pi,\pi]$ and the corresponding energies $\bs{y}_{\mathrm{t}}$ and $\bs{y}_{\mathrm{v}}$ are evaluated noiselessly on a classical computer. In both plots the horizontal axis is the size of the training data set $N_{\mathrm{t}}$, vertical axis is the validation score (defined in \eqref{eq:val_score}). \textbf{a.)} Results for approximated state kernels with added noise terms and the full un-approximated state kernel. \textbf{b.)} Results for the full unitary kernel and various classical kernels.
    In both plots we show the median over 100 repeats (solid lines/dots) and the inter-quartile range of the data (filled). Maximum likelihood hyperparameter optimization with respect to the training data was used to optimize any kernel hyperparameters (including the strength $\sigma_n^2$ of the noise terms added to approximated state kernels).}
    \label{fig:mps_hybrid_plot}
\end{figure*}

Figure \ref{fig:mps_hybrid_plot}a shows the results of a repeat of the simulations in \ref{fig:mps_plot}a when the approximated kernels are equipped with this additional noise term. For comparison with the classical kernels used in our other simulations, the plot in Figure \ref{fig:mps_hybrid_plot}b shows the validation scores obtained for GP models using these classical kernels as well as the unitary kernel. Because the $U_B$ used in these simulations only has a depth of 1 (a single layer of entangling gates) the unitary kernel can be calculated classically at low cost and so provides a benchmark against which the approximated state kernels should be compared.

When given an additional noise term, the approximated state kernels no longer suffer from an eventual drop in predictive accuracy due to feature space saturation. Instead, their performance continues to improve as more data is provided until the validation score eventually begins to saturate. For $\chi_{\mathrm{max}}>1$ the approximated state kernels all yield better performance than any of the classical kernels and the unitary kernel. Although the unitary kernel fares better than the classical kernels, its performance increases slowly with the number of training data due to it having a much larger feature space dimension than any of the (approximated) state kernels. Again we see for small amounts of data, the state kernel approximations with low $\chi_{\mathrm{max}}$ have higher validation scores than the full state kernel, particularly with $\chi_{\mathrm{max}}=2$ for small $N_{\mathrm{t}}$ and $\chi_{\mathrm{max}}=4$ for intermediate $N_{\mathrm{t}}$. This reinforces the idea that having a simpler model (with a smaller feature space) may provide accuracy benefits in the small-data regime as well as being easier to calculate. This raises the possibility that one could adaptively switch between kernels of varying complexity as more data is obtained, using comparative tools like the Bayes factor to decide which is most suitable at the current optimization stage. 

\section{Analysis of quantum kernel feature spaces}\label{sec:feat_construction}
Many of our numerical results can be understood by analysing the dimensions of the feature spaces induced by the state and unitary kernels.
From \eqref{eq:state_fs}, it is immediately apparent that the state kernel's feature space $\mathcal{F}_s$ is finite-dimensional, unlike the classical kernels we have considered which have infinite dimensional \cite{williams2006} feature spaces. This has the consequence that this kernel is \emph{not universal} meaning that it cannot be used to approximate an unknown (well-behaved) function to an arbitrary degree of accuracy \cite{scholkopf2001}. The feature map sends an input vector of $p$ gate angles to vector with at most $3^p$ elements (which are linear sums of various Fourier components). Using the identity $A\otimes C^T\kket{B}=\kket{ABC}$ one can identify the set of (unnormalized) states $\{s_i \kket{\rho_0}\}$ whose inner products form $\bs{S}$ as a set of vectorized Hermitian operators on $\mathcal{H}$. As there can only be at most $d^2$ linearly independent Hermitian operators in a $d$-dimensional Hilbert space this puts an additional upper bound on the state kernel's feature space dimension. The maximal dimension for an $n$-qubit, $p$-parameter state kernel's feature space is therefore $\min(3^p, 4^n)$. In practice, the dimension of $\mathcal{F}_s$ may be lower than this due to redundancies or constraints on the set of Pauli rotations used in the ansatz (which reduces the span of the set of vectorized operators spanned by $\bs{s}\kket{\rho_0}$ and so the rank of $\bs{S}$), or correlations between parameters (which would reduce the number of linearly independent components of $\bs{v}(\ptheta)$). 

Assuming no ansatz redundancies, this implies that for an $n$-qubit ansatz we can hit the upper bound for $\dim{\mathcal{F}_s}$ imposed by the dimension of $\mathcal{H}$ (i.e. $3^p>4^n$) using  $\ceil{\frac{2}{\log_2 3} n}\approx\ceil{1.262 n}$ PPRs. However, in Appendix \ref{appendix:fs_s} we show that in some cases the feature space dimension is also limited by constraints on the linear independence of the elements of $\bs{s}\kket{\rho}$, meaning that more than $\ceil{\frac{2}{\log_2 3} n}$ parameterized rotations are required. We derive the following recursion relation for the maximal dimension $d_s^{(n,p)}$ of $\mathcal{F}_s$ for an $n$-qubit circuit with arbitrary fixed unitaries and $p$ parameterized Pauli rotations is given by
\begin{equation}\label{eq:state_scaling}
    d_s^{(n,p)}\coloneqq\min\left(4^n, 3d_s^{(n,p-1)}, \frac{4^n}{2} + d_s^{(n,p-1)}\right),
\end{equation}
where for completeness we define the base-case $d_s^{(n,0)}=1$ for all $n$. By finding $p$ such that the $p+1^{\mathrm{th}}$ rotation first hits the linear independence bound, i.e. when $3^{p+1}>\frac{4^n}{2}+3^p$, we find that total required number of parameterized rotations to obtain a maximal dimension feature space is equal to the upper bound $\ceil{\frac{2}{\log_2 3} n}$ if $\ceil{\frac{2}{\log_2 3} n-\frac{1}{\log_2 3}}=\ceil{\frac{2}{\log_2 3} n}-1$, or $\ceil{\frac{2}{\log_2 3} n}+1$ if $\ceil{\frac{2}{\log_2 3} n-\frac{1}{\log_2 3}}=\ceil{\frac{2}{\log_2 3} n}$.

The results shown in Figure \ref{fig:gp_full} give evidence of additional ansatz-imposed constraints on the state kernel's feature space dimension. With $16$ parameters in the ansatz and $4$ qubits, the absolute upper bound on the feature space dimension and thereby the number of training data points for $R_{\mathrm{v}}^2=1$ should be $4^{4}=256$ however this instead occurs with just $136$ observations. The ansatz used for these simulations (shown in Fig. \ref{fig:ansatz}) creates wavefunctions with real-valued amplitudes (in the computational basis), restricting the state space it can explore. The states it can produce have density matrices with components containing only even numbers of $Y$ Pauli operators (and are generated by Pauli rotations with $Y$s acting on odd numbers of qubits) \cite{tang2021}. By counting the number of real $n$-qubit Pauli operators we can obtain a tighter upper bound for the maximal feature space dimension of real ansatz circuits. As there are $3^{n-2j}$ combinations of $\{I,X,Z\}$ for all ${n \choose 2j}$ placements of $2j$ $Y$s, this upper bound is given by $\sum_{j=0}^{\lfloor n/2\rfloor}3^{n-2j}{n\choose 2j}$ which for $n=4$ gives a dimension of $136$. This was confirmed by generating Gram matrices with this kernel for $>136$ uniform randomly chosen points and calculating their rank (and so the dimension of the kernel), again yielding a dimension of $136$.

We also note that $I\otimes I\pm P\otimes P^*$ is a projection onto the $\pm1$ eigenspace of the Pauli operator $P\otimes P^*$ while $I\otimes iP^* - iP\otimes I=I\otimes iP^*(I\otimes I - P\otimes P^*)$ is a projection onto the $-1$ eigenspace of $P\otimes P^*$ followed by a (Clifford) unitary $I\otimes iP^*$. This means that if the initial state is a stabilizer state (such as the usual computational basis state $\ket{0\dots0}$) and all $R_1,\dots, R_p$ are Clifford operations, the elements of $\bs{s}\kket{\rho_0}$ and the overlaps in $\bs{S}$ can be calculated efficiently on a classical computer as a $2n$-qubit stabilier simulation \citep{Garcia2012,Aaronson_2004}. An obvious limitation however is that the feature space dimension scales exponentially in the number of parameterized rotations. This is not at all surprising as an initial stabilizer state with only a few non-Clifford rotations applied can be simulated ``efficiently" (i.e. at cost polynomial in the number of qubits $n$, but exponential in the number of non-Clifford gates) on a classical computer \cite{Bravyi2019}.


The feature map for the unitary kernel $\bs{\varphi}_u$ is qualitatively similar to $\bs{\varphi}_s$, mapping input vectors into a finite-dimensional feature space vectors whose elements are sums of Fourier components. As $\bs{v}(\ptheta)$ makes a reappearance here we again have an upper-bound to $\dim{\mathcal{F}_u}$ of $3^p$ for a $k$-Pauli rotation ansatz. We show in appendix \ref{appendix:fs_u} that there is a further upper bound due to the finite dimensionality of the $n$ qubits' Hilbert space of $4^{2n}-2(4^n-1)$ meaning that the overall maximal feature space dimension for the unitary kernel with $n$ qubits and $p$ PPRs is 
\begin{equation}\label{eq:unitary_scaling}
    d_u^{(n,p)} \coloneqq \min\left(4^{2n}-2(4^n-1), 3^p\right).
\end{equation}

Broadly speaking, the global predictive accuracy of a kernel regression model depends on whether the objective function is linear in the kernel's feature space and, for finite-dimensional kernels, the fraction of the kernel's feature space spanned by the observed data \cite{williams2006}. If the (assumed noiseless) data fully spans the kernel feature space then any new point can be written as a linear combination of the feature mapped observations, meaning a GP model based on the kernel will be perfectly accurate. For both the state and unitary kernels the linearity condition holds (in the absence of noise) however the key difference between them is in the scaling of their feature spaces. In most practical applications an ansatz circuit will have a number of parameterized gates polynomial in the number of qubits $p\sim\text{poly}(n)$ -- \eqref{eq:state_scaling} suggests that this would typically lead to a $\mathcal{F}_s$ of maximal dimension of $4^n$ (although this may be reduced by ansatz redundancies or specific circuit structure). To build a globally accurate (high predictive accuracy for all $\ptheta$) state kernel-based regression model for the energy of an arbitrary $n$-qubit Hamiltonian we would need $\mathcal{O}(4^n)$ training data. In contrast, the unitary kernel's maximal feature space dimension scales like $O(4^{2n})$ and so one would require roughly quadratically more data to achieve the same global accuracy. This is less of an issue when considering local regression, which is equally important in Bayesian optimization, where the data only needs to span a subspace of the feature space local to the point of interest. This can be seen particularly in \ref{fig:16_prm_local}b in which the unitary kernel and classical kernels compete relatively well with the state kernel in highly localized GP regression.

\section{Conclusion}
The framework we have presented here allows on-device VQE of small systems to be performed with remarkably few energy evaluations. 
Our use of a Gaussian process surrogate model equipped with a classically-evaluated quantum kernel allows us to avoid many of the difficulties faced when implementing gradient-based optimization on NISQ devices.
The two quantum kernels we have considered are based on the similarity (in terms of the fidelity) between two parameterized quantum states and two unitary operations. 
We have demonstrated that these kernels can be used to build very powerful Gaussian process surrogate models which are manifestly well-suited for regressing the cost function in VQE.
For this regression task, these quantum kernel-based surrogate models exhibit significantly better predictive accuracy over many widely-used classical kernels.
The advantage in predictive accuracy is particularly acute for the state fidelity-based kernel.
Using this kernel one can build accurate regression models with far fewer samples than is needed by the unitary kernel or the classical kernels we have considered.

The advantage brought by a state kernel-based GP surrogate holds both on local and global parameter scales. Accuracy on these scales determines the effectiveness of the exploitative and exploratory phases of Bayesian optimization. Therefore, these results suggest that a state kernel-based GP surrogate can allow for fast Bayesian VQE with few energy evaluations. 
Due to the large number of kernel evaluations needed to build and optimize a quantum kernel based GP surrogate, it is impractical to evaluate these kernels on-device. 
Instead we propose that VQE problems at a scales where quantum kernel evaluation remains classically tractable can be solved quickly and with high accuracy using classically-evaluated quantum kernels. 
This allows one to make use of prior knowledge of the ansatz circuit while implicitly obtaining and making use of knowledge about a device's noise processes through the observed energy values. 
Through numerical experiments we have verified that this Bayesian approach to VQE is effective with both noiseless and realistic noisy energy observations.

The suitability of the state kernel for VQE stems from its induced feature space. We have explicitly constructed feature maps for the state and unitary kernels and demonstrated that the energy function in VQE is linear in both feature spaces. The dimension of the state kernel is roughly quadratically smaller than that of the unitary kernel (for typical ansatz circuits) and we believe this is the origin of the state kernel's advantage over the unitary kernel in GP regression and Bayesian optimization of VQE.

Finally, while we have shown our method for VQE is highly effective on small numbers of qubits, classical evaluation of the quantum kernels for more qubits quickly becomes intractable.
By letting only individual blocks/layers of parameterized gates vary at any one time one can ensure that the unitary kernel remains tractable even for large numbers of qubits. 
For the state kernel, which is more useful for VQE, as the input state before varied gates must be considered to evaluate the kernel which may be infeasible. 
To remedy this we presented a scheme for approximating the state kernel so that it may be evaluated classically for large numbers of qubits. 
We ensure that the computational cost of evaluating this approximated state kernel remains bounded by approximating the input state to a matrix product state with a limited bond-dimension. 
Our simulations suggest that these approximated state kernels exhibit similar performance to the full state kernel in global GP regression provided the maximum allowed bond dimension is sufficiently large. Our results also suggest that GP models using an approximated state kernel with a lower maximum bond dimension (and so a smaller feature space dimension) can outperform the full state kernel in situations where few energy observations have been made. 
\begin{acknowledgments}
\textbf{Acknowledgements:}
We acknowledge the use of IBM Quantum services for this work. The views expressed are those of the authors, and do not reflect the official policy or position of IBM or the IBM Quantum team. 
\textbf{Funding:}
This work is supported by the Samsung GRC grant and the UK Hub in Quantum Computing and Simulation, part of the UK National Quantum Technologies Programme with funding from UKRI EPSRC grant EP/T001062/1.
\end{acknowledgments}
\section{Methods}
\subsection{Gaussian processes}\label{sec:GPs}
The kernel-based surrogate model used in our Bayesian VQE simulations is a Gaussian process. A Gaussian process (GP) is a collection of random variables with the property that any finite subset has a joint multivariate normal distribution \cite{williams2006}. It is defined in terms of a covariance (a kernel) function $k(\cdot,\cdot)$ and a mean function $\mu(\cdot)$. Given a vector of $m$ observed values $\bs{y}=(y_1,\dots,y_m)^T$ of some unknown function at points $\mathbf{X}=(\bs{x}_1,\dots, \bs{x}_m)^T$ (expressed as an $m\times p$ matrix, where $p$ in the input space dimension) then the joint distribution with a new point $y^*$ at location $\bs{x}^*$ is \cite{williams2006}
\begin{equation}
    \begin{bmatrix}
        \ \bs{y}\ \\
        \ y^*
    \end{bmatrix}\sim\mathcal{N}\left(
        \begin{bmatrix}
        \ \bs{\mu}\ \\
        \ \mu(\bs{x}^*)
    \end{bmatrix},
    \begin{bmatrix}
        \mathbf{K}&\bs{k}\\
        \bs{k}^T&k(\bs{x}^*,\bs{x}^*)
    \end{bmatrix}
    \right),
\end{equation}
where $\bs{\mu}$ is an $m$ element  mean vector with $\mu_i=\mu(\bs{x}_i)$, $\mathbf{K}$ is an $m\times m$ Gram matrix of pair-wise kernel evaluations between the observed inputs, $K_{ij}=k(\bs{x}_i,\bs{x}_j)$, and $\bs{k}$ is an $m$ element vector of kernel evaluations between the new point $\bs{x}^*$ and the observed inputs $\mathbf{X}$, $k_i=k(\bs{x}^*,\bs{x}_i)$. If noise is present in the observations this is usually approximated to be normally distributed with zero mean. One can show analytically that for observations with normally distributed noise the kernel should be altered to $k(x_i,x_j)\to k(x_i,x_j)+\sigma_n^2\delta_{ij}$, where $\sigma_n^2$ is a noise variance hyperparameter and the indices of $\delta_{ij}$ correspond to \tcr{indices} in the training data meaning that $\mathbf{K}\to \mathbf{K}+\sigma_n^2 I$ but $\bs{k}$ and $k(\bs{x}^*,\bs{x}^*)$ are unchanged. Marginalising over the observed data, the resulting posterior distribution is itself a normal distribution with 
a posterior mean (a prediction) $\mathbb{E}[y^*|\bs{x}^*,\mathbf{X},\bs{y}]=\hat{y}(\bs{x}^*)$ given by
\begin{equation}\label{eq:gp_mean}
    \hat{y}(\bs{x}^*)=\mu(\bs{x}^*) + \bs{k}^T(\mathbf{K}+\sigma_n^2I)^{-1}(\bs{y}-\bs{\mu})
\end{equation}
and a posterior variance $\mathrm{Var}[y^*(\bs{x}^*)]={\Delta y(\bs{x}^*)}^2$ of
\begin{equation}\label{eq:gp_var}
    {\Delta y(\bs{x}^*)}^2=k(\bs{x}^*,\bs{x}^*) - \bs{k}^T(\mathbf{K}+\sigma_n^2I)^{-1}\bs{k}.
\end{equation}

The performance of GP regression depends on the suitability of the kernel function in describing the unknown function. Unless one has significant prior knowledge of the problem at hand, choosing a suitable kernel is often difficult. As a result, problem-agnostic approaches to GP regression often make use of flexible kernels with internal hyperparameters which are varied to fit observed data. A common approach is to find hyperparameter values that maximise the marginal likelihood of the observed data (often the log-likelihood) \cite{williams2006}. \tcr{Thanks} to the normality of the GP's distributions, \tcr{the} marginal log-likelihood $M(\mathbf{X},\bs{y})=\log{p(\bs{y}|\mathbf{X},\alpha)}$ can be calculated analytically for a kernel with hyperparameter $\alpha$ as 
\begin{widetext}
\begin{equation}\label{eq:ml}
    M(\mathbf{X},\bs{y})=-\frac{1}{2}(\bs{y}-\bs{\mu})^T(\mathbf{K}+\sigma_n^2\mathbf{I})^{-1}(\bs{y}-\bs{\mu})-\frac{1}{2}\log \det[\mathbf{K}+\sigma_n^2\mathbf{I}]-\frac{n}{2}\log 2\pi.
\end{equation}
\end{widetext}
The gradient of this quantity with respect to a kernel hyperparameter $\alpha$ is given by
\begin{equation}\label{eq:gradml}
\begin{split}
    \frac{\partial M(\mathbf{X},\bs{y})}{\partial \alpha}=\frac{1}{2}\Tr{(\bs{\beta}\bs{\beta}^{T}-\mathbf{K}^{-1}_{\sigma_n})\frac{\partial\mathbf{K}_{\sigma_n}}{\partial\alpha}},\\\text{where } \bs{\beta}=\mathbf{K}^{-1}_{\sigma_n}\bs{y},\ \mathbf{K}_{\sigma_n}=\mathbf{K}+\sigma_n^2\mathbf{I},   
\end{split}
\end{equation}
and so this quantity can be maximised through gradient ascent at a cost primarily dictated by the calculation of $K_{\sigma_n}^{-1}$ if the kernel evaluations and their gradients are inexpensive to calculate. 

The posterior variance \eqref{eq:gp_var} depends only on kernel evaluations and not the observations $\bs{y}$. If the kernel is fixed, a re-scaling of the training data of the problem $\bs{y}\to\alpha\bs{y}$ will leave the posterior variance unchanged whereas it should be a factor of $\alpha^2$ larger. 
To allow the posterior variance to scale with the size of the observations an external ``signal variance" hyperparameter $\sigma^2$ is often added to the kernel $k(\bs{x},\bs{x}')\to\sigma^2 k(\bs{x},\bs{x}')$. We see from \eqref{eq:gp_mean} and \eqref{eq:gp_var} that this does not change posterior mean prediction (up to a re-scaling of $\sigma_n^2$) but the posterior variance is scaled by $\sigma^2$. If the base kernel has no hyperparameters (which is true for the quantum kernels we consider) and the observed data is assumed to be noiseless then it is simple to show the marginal likelihood of the training data is maximised when $\sigma^2_{\mathrm{noiseless}}=(\bs{y}-\bs{\mu})\mathbf{K}^{-1}(\bs{y}-\bs{\mu})^T/m$ (where $\mathbf{K}$ is the Gram matrix for the kernel without $\sigma^2$ and $m$ is the number of observations) -- this often serves as a good starting point for maximum likelihood optimization even if noise is present.



The mean function $\mu$ is often set to a constant value, most commonly $\mu(\bs{x})=0$ for all $\bs{x}$, as it is primarily the covariance function which defines a GP's properties; in most cases $\mu$ only has a significant impact at points that are far away (with respect to the kernel) from any observed data. However (as shown in appendix \ref{sec:feat_construction}) for many ansatz circuits the expected value (taken over the gate angles) of the noiseless energy $\mathbb{E}_{\ptheta}[E(\ptheta)]$ is given by $\Tr{H}$ meaning that one should set $\mu(\bs{x})=\Tr{H}$. Alternatively, one can remove this diagonal offset from the Hamiltonian by using $H' = H-\Tr{H}I$ and set $\mu=0$. We suggest that this should be done generally as calculating or estimating the average energy for a complicated ansatz will be difficult, especially if device noise is present. Removing this offset is good practice in general as a large diagonal offset in $H$ can lead to improper conclusions on the effectiveness of an optimization scheme. 

In our simulations we build the GP models using two quantum kernels (described in section \ref{sec:qkernels}) and a set of classical kernel functions. These classical kernels are as follows; Matern kernels with smoothness hyperparameter $\nu=5/2$ and $\nu=3/2$, the radial basis function (RBF) kernel, and the rational quadratic (RQ) kernel. Their kernel functions are:
\begin{widetext}
\begin{equation}
    k_{Matern}(\bs{x},\bs{x}';\nu=5/2,l ) = \left(1+\frac{\sqrt{5}}{l}d(\bs{x},\bs{x}')+\frac{5}{3l}d(\bs{x},\bs{x}')^2\right)\exp{-\frac{\sqrt{5}}{l}d(\bs{x},\bs{x}')},
\end{equation}
\begin{equation}
    k_{Matern}(\bs{x},\bs{x}';\nu=3/2,l ) = \left(1+\frac{\sqrt{3}}{l}d(\bs{x},\bs{x}')\right)\exp{-\frac{\sqrt{3}}{l}d(\bs{x},\bs{x}')},
\end{equation}
\begin{equation}
    k_{RBF}(\bs{x},\bs{x}';l )=\exp{-\frac{d(\bs{x},\bs{x}')^2}{2l^2}},\text{ and}
\end{equation}
\begin{equation}
    k_{RQ}(\bs{x},\bs{x}';l,\alpha)=\left(1+\frac{d(\bs{x},\bs{x}')^2}{2\alpha l^2}\right)^{-\alpha},
\end{equation}
\end{widetext}
where $d(\bs{x},\bs{x}')$ is the squared euclidean distance between points $\bs{x}$ and $\bs{x}'$, $l$ are length scale hyperparameters, and $\alpha$ is a scale mixture hyperparameter for the RQ kernel.

These classical kernels require optimization of internal hyperparameters (a length scale $l$ and, for the RQ kernel, a mixing parameter $\alpha$) to ensure that the kernel's infinite-dimensional feature space most-succinctly describes the objective function. By contrast, the state and unitary kernel have finite-dimensional feature spaces in which the (noiseless) energy function is manifestly linear which means no hyperparameter optimization is required for a GP model to make sensible predictions about the energy. Hyperparameter optimization is often the most computationally intensive part of Bayesian optimization as it requires repeated calculation of the Gram matrix $\mathbf{K}$ (see \eqref{eq:ml} and  \eqref{eq:gradml}) and its inverse. If we have a kernel with no hyperparameters other than those for the signal variance and noise $k(\bs{x}_i,\bs{x}_j)=\sigma^2k_0(\bs{x}_i,\bs{x}_j)+\sigma_n^2\delta_{ij}$, where $k_0$ is fixed, then the Gram matrix for the fixed part of the kernel $\mathbf{K}_0$ only needs to be calculated once. This simplifies the hyperparameter optimization as the total Gram matrix $\mathbf{K}=\sigma^2 \mathbf{K}_0 + \sigma_n^2 I$ can be calculated easily for $\sigma^2$ and $\sigma_n^2$ from $\mathbf{K}_0$. Additionally, the eigenvalues/vectors of the total $\mathbf{K}$ can be found analytically in terms of those for $\mathbf{K}_0$, simplifying calculation of the inverse Gram matrices needed to both optimize the hyperparameters and make predictions from the model. This ultimately means that the state and unitary kernels only need to be evaluated once between any two data points within a training data (up to a re-scaling by $\sigma^2$ and a noise offset), in contrast to the classical kernels whose Gram matrices must be entirely recalculated whenever an internal hyperparameter is adjusted.

\subsection{Bayesian optimization}\label{sec:BO}
In Bayesian optimization one attempts to minimize an unknown expensive-to-evaluate objective function by assuming it is randomly drawn from some family of functions \cite{jones1998}. A prior distribution is chosen to encode any initial beliefs about this function, usually that it is sampled from a Gaussian process. From observations of the function's values and the prior, a posterior distribution over the chosen family of functions is constructed (the posterior distribution of the GP), quantifying how likely it is that a given function could have produced the observations.
An acquisition function is then calculated from the posterior distribution and is maximised/minimised to determine the next point to query. 
Once the new point has been queried, the posterior distribution is reconstructed with this new data and the process is repeated until convergence to the objective function global minima.

The acquisition function quantifies how ``promising" the querying of a given an unseen point appears for minimizing the objective function. These functions are often designed to balance the exploitation of regions of the parameter space that have already shown good objective function values with the exploration of areas in the parameter space where observations are sparse. By only querying parameter points that appear likely to yield a significant improvement in the objective function, Bayesian optimization can allow global minima to be found with remarkably few iterations. In our simulations we will use the Expected Improvement acquisition function \cite{jones1998} given (for minimization) by $\mathrm{EI}(\bs{x})=\mathbb{E}[\mathrm{max}(y_{\mathrm{best}}-\hat{y}(\bs{x})+\xi,0)|\bs{x}^*,\mathbf{X},\bs{y}]$, where the expectation is taken over the surrogate model's posterior distribution, $y_{\mathrm{best}}=\min\{\bs{y}\}$ is the lowest observation seen so far, and $\xi$ is an exploration hyperparameter. Using of the analytically tractable GP posterior distribution (i.e. normally distributed with mean/variance given by \eqref{eq:gp_mean} and \eqref{eq:gp_var}) we can derive an explicit formula for this \cite{jones1998}:
\begin{equation}\label{eq:EI}
\begin{split}
    \mathrm{EI}(\bs{x})&\coloneqq(y_\mathrm{best}-\hat{y}(\bs{x})+\xi)\Phi(Z) + \Delta y(\bs{x})\phi(Z),\\&\text{ where } Z=\frac{y_\mathrm{best}-\hat{y}(\bs{x})+\xi}{\Delta y(\bs{x})},
\end{split}
\end{equation}
$\Phi$ and $\phi$ are the cumulative distribution and probability density functions for the standard normal distribution respectively. Expected improvement-based optimization can sometimes be overly greedy and exploitative and so a hyperparameter $\xi$ has been added to control the ratio of exploration to exploitation. The $\xi$ parameter allows more positive observations (up to where  $y^*(\bs{x})=y_{\mathrm{best}}+\xi$) to count as an improvement, meaning the objective function encourages exploration further away from the point where $y_{\mathrm{best}}$ was observed. In our simulations we used $\xi=0.01$ and optimized the surrogate model using the L-BFGS-B gradient-descent algorithm \cite{byrd1995}.

While it is possible to run Bayesian optimization starting with an observation at a single randomly chosen parameter point, more typically a collection of initial points are used to ensure an initial explorative phase. While there are many designs for these points \cite{Bossek2020} we use the simplest; random uniform sampling across the parameter space.

\subsection{MPS approximation to the state kernel}\label{sec:mps}
Our approximation scheme for the state kernel relies on the parameterized circuit being of the form $\ket{\psi(\ptheta_B)}=U_A(\ptheta_A)U(\ptheta_B)U(\ptheta_C)\ket{\bs{0}}$ where $\ptheta_A$ and $\ptheta_C$ are fixed, so that the state kernel can be written in the form \eqref{eq:limited_skern}. We also require that $U_B$ is simple enough that its action on an MPS state of relatively low bond dimension can be simulated classically. If this is the case then by approximating $\ket{\psi_A}$ to an MPS $\ket{\tilde{\psi}_A^{(\chi_\mathrm{max})}}$, we can achieve a classically tractable approximation to the state kernel. 
\begin{figure*}
    \centering
    \includegraphics[width=0.7\linewidth]{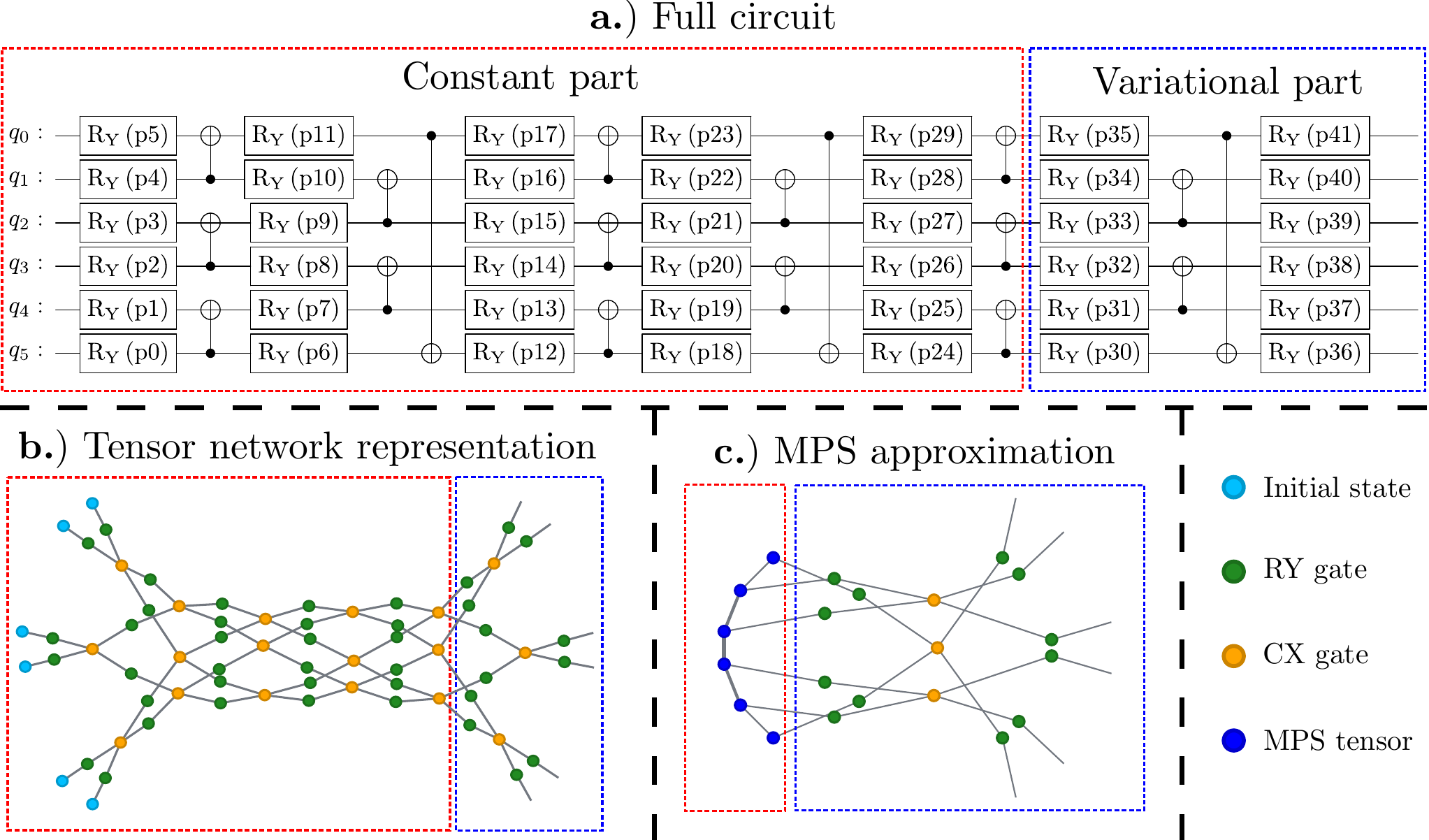}
    \caption{\textbf{Illustration of the scheme used to produce an MPS approximation to the state kernel.} We assume the ansatz circuit has a limited number of parameterized gates as would be used in layer/block-wise VQE. \textbf{a.)} The portions of the ansatz circuit which are relevant to the state kernel are separated into a constant part $U_A$ and a parameterized part $U_B$. \textbf{b.)} The anastz circuit is re-expressed in a tensor network representation. \textbf{c.)} The input state to $U_A$ is converted to an MPS state with bond dimension $\chi=1$ into which the gates are sequentially contracted (using the contraction and truncation process illustrated in Figure \ref{fig:tn_diag}).}
    \label{fig:mps_approximation_scheme}
\end{figure*}

Figure \ref{fig:mps_approximation_scheme} illustrates our approximation scheme. We begin (\ref{fig:mps_approximation_scheme}a) with a quantum circuit composed of a constant part ($U_A$) and a variational part ($U_B$) whose parameters are to be optimized. For simplicity we omit any $U_C$ that lies in the future light-cone of the variational part of the circuit as this has no effect on the state kernel. The circuit is then converted into a tensor-network representation, which can be done at a cost proportional to the number of gates (an example is shown in \ref{fig:mps_approximation_scheme}b). The initial state $\ket{\bs{0}}$ is then converted to a matrix product state with bond dimension $\chi=1$ and each gate in the circuit is contracted into the MPS to produce $\ket{\tilde{\psi}^{(\chi_\mathrm{max})}_A}$ (shown in \ref{fig:mps_approximation_scheme}c).

\begin{figure*}
    \centering
    \includegraphics[scale=0.45]{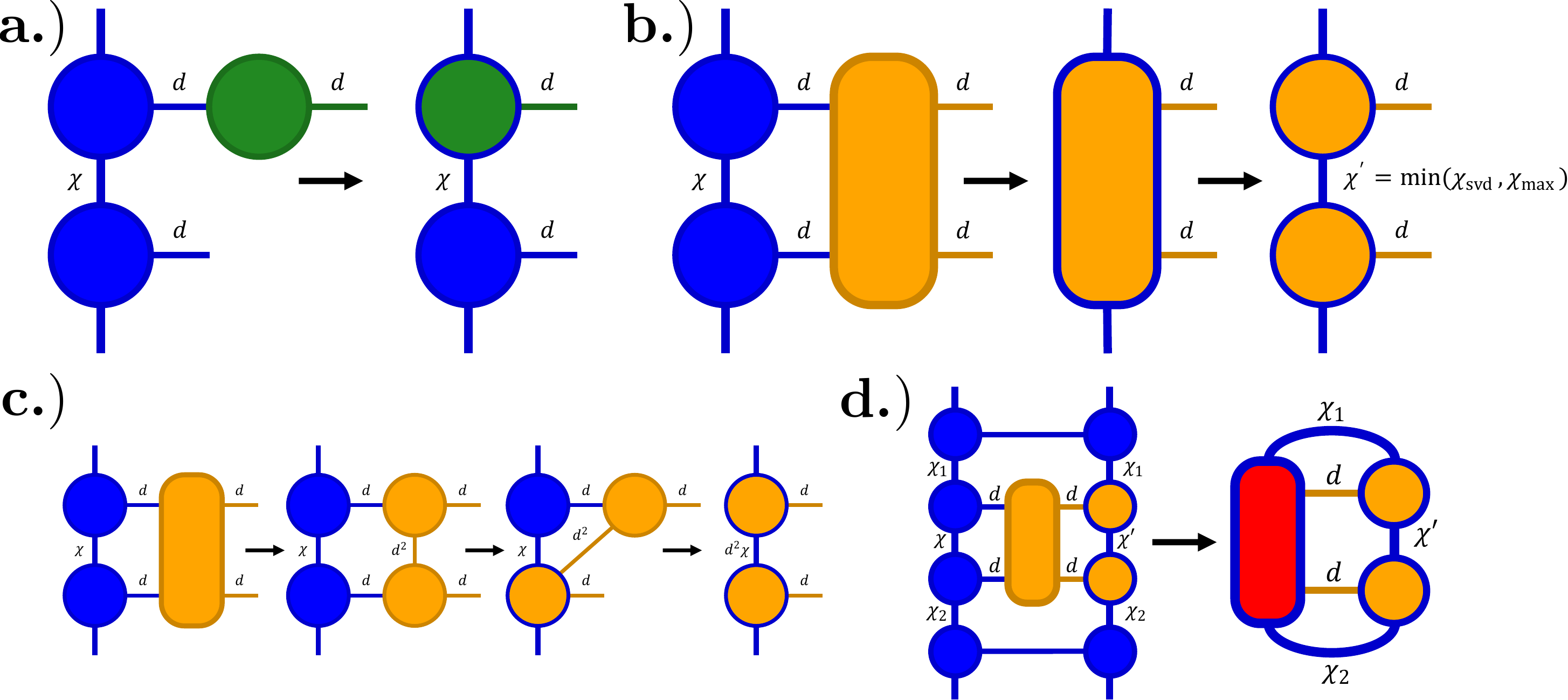}
    \caption{\textbf{Illustration of process for contracting single- and two-qubit gates into a matrix product state.} The initial matrix product state is shown as a series of connected MPS tensors (blue circles with dark blue outlines), one at each qubit/site. Only relevant sites are shown in each diagram. Relevant bond dimensions are shown; $\chi$ denotes an internal bond while $d$ denotes $d$-dimensional physical bonds ($d=2$ for qubits). \textbf{a.)} A single qubit gate is applied by contracting its unitary (shown as a green dot) directly into the MPS tensor of the corresponding qubit, leaving the MPS structure and bond dimensions unchanged. \textbf{b.)} A two-qubit gate on adjacent qubits is applied by contracting its unitary (yellow oblong with dark yellow borders) into the MPS tensors of the corresponding qubits and splitting the result with SVD.
    The bond dimension is capped to a value $\chi'=\min(\chi_{\mathrm{svd}}\chi_\mathrm{max})$ by retaining only the $\chi_\mathrm{max}$ largest singular values, after which the state is renormalized. \textbf{c.)} An alternative two-qubit gate application scheme demonstrating that the maximal value of $\chi'$ is $d^2\chi$.
    \textbf{d.)} Following truncation the fidelity between the produced MPS state and the non-truncated state (modulus squared of the tensor network shown) is optimized to improve accuracy. 
    This optimization is similified by only varying the tensors altered when applying the gate and by pre-contracting all other tensors to form the red tensor shown.}
    \label{fig:tn_diag}
\end{figure*}

Figure \ref{fig:tn_diag} illustrates the process of contracting a gate $G$ into an MPS $\ket{\psi_\mathrm{MPS}}$ (with physical dimension $d$). As we only consider single- and two-qubit gates (outlined in green and yellow respectively) only two MPS tensors with shared bond dimension $\chi$ are shown (outlined in blue). Vertical open bonds indicate connections to the rest of the MPS state. In \ref{fig:tn_diag}a a single qubit unitary is contracted directly into the MPS tensor at the corresponding site/qubit, altering this tensor but leaving the bond dimensions unchanged. Figure \ref{fig:tn_diag}b shows the contraction of a two-qubit unitary which acts on adjacent sites. The two-qubit unitary is fully contracted into the MPS tensors yielding a joint tensor for the two sites. Singular value decomposition (SVD) is used to split this joint tensor into two new MPS tensors with a new bond dimension $\chi'$. The size of $\chi'$ depends on the specifics of the unitary and the state to which it is applied but takes a maximum value of $d^2\chi$. Figure \ref{fig:tn_diag}c shows this diagrammatically by giving an alternative scheme for contracting the two-qubit gate. Here the unitary is first decomposed using SVD into two tensors connected by a bond of dimension $d^2$. Each resulting tensor is contracted into their connected MPS tensor yielding an MPS state with a two bonds between the new MPS tensors (of size $\chi$ and $d^2$). These bonds are combined to form a single bond of dimension $d^2\chi$. Two qubit gates on non-neighbouring sites can be contracted using the same scheme by applying a chain of SWAP gates (as described in \cite{Vidal2003}) but can significantly increase the bond dimensions.

Applying many two-qubit gates will lead to exponential growth in the MPS's bond dimensions. To avoid this we truncate $\chi'$ by only retaining the $\chi_\mathrm{max}$ largest singular values obtained from the SVD in the final stage of \ref{fig:tn_diag}b. Following truncation the MPS is re-normalized, yielding an approximation $\ket{\tilde{\psi}_{MPS}}$ to the transformed state $G\ket{\psi_\mathrm{MPS}}$ with a maximum bond dimension of $\chi_{\mathrm{max}}$. If significant \tcr{singular} values of the joint tensor are discarded, the fidelity between $\ket{\tilde{\psi}_{MPS}}$ and the target state $G\ket{\psi_\mathrm{MPS}}$ will be degraded. We partially mitigate this by maximizing the fidelity $\abs{\bra{\tilde{\psi}_{\mathrm{MPS}}}G\ket{\psi_\mathrm{MPS}}}^2$ between these two states, optimizing over $ \ket{\tilde{\psi}_{\mathrm{MPS}}}$. This procedure can be greatly simplified by only optimizing over the two MPS that are altered when producing $\ket{\tilde{\psi}_{MPS}}$ (shown as yellow dots with a blue outline in Figure \ref{fig:tn_diag}b and \ref{fig:tn_diag}d). By first contracting over the tensors which are not optimized (creating the tensor shown with red fill Figure \ref{fig:tn_diag}d) this optimization can be performed extremely efficiently, involving contractions over just 4 tensors. These initial contractions can also be further simplified by first canonicalizing $\ket{\psi_\mathrm{MPS}}$ around the two qubits acted upon by the unitary \cite{Bridgeman2017} (provided the MPS has open boundaries to allow this). Once all gates in the circuit have been contracted into the MPS using the scheme we have outlined (with the necessary truncation and fidelity optimization) the resulting MPS state $\ket{\tilde{\psi}_A^{(\chi_\mathrm{max})}}$ is used as the input state for the approximated state kernel given in \eqref{eq:limited_skern}.

\clearpage
\bibliographystyle{unsrt}
\bibliography{refs}
\clearpage
\appendix
\onecolumngrid
\section{Derivation of feature spaces}\label{appendix:feature_spaces}
The state kernel for the ansatz in \eqref{eq:pauli_ansatz} is given by:
\begin{equation}
\begin{split}
    k_s(\bs{\theta},\bs{\theta}')&=\Tr{U(\bs{\theta}')\rho_0 U^\dagger(\bs{\theta}')U(\bs{\theta})\rho_0U^\dagger(\bs{\theta})}\\
    &=\bbrakket{U(\bs{\theta}')\rho_0 U^\dagger(\bs{\theta}')}{U(\bs{\theta})\rho_0U^\dagger(\bs{\theta})}\\
    &=\bbra{\rho_0}(U^\dagger(\ptheta')\otimes U^T(\ptheta'))(U(\ptheta)\otimes U^{*}(\ptheta))\kket{\rho_0},
\end{split}
\end{equation}
where $\rho_0=\ket{\bs{0}}\bra{\bs{0}}$, $\kket{A}$ denotes the vectorization of operator $A$ and we have made use of the identities $\bbrakket{A}{B}=\Tr{A^\dagger B}$ and $\kket{ABC}=A\otimes C^{T}\kket{B}$. At this point, we could stop and identify the state kernel's feature vectors as quantum states on $\mathcal{H}\otimes\mathcal{H}^*$ (where $\mathcal{H}$ is the Hilbert space of the qubits) of the form $U(\ptheta)\otimes U^{*}(\ptheta)\kket{\rho_0}$. However in doing so would have to construct these states explicitly, making analysis and comparison to classical kernels difficult. 
By expanding each Pauli rotation as $e^{-iP_i\theta_i/2}=\cos(\frac{\theta_i}{2})I - i\sin(\frac{\theta_i}{2})P_i$ we can express $U(\ptheta)$ in a convenient form as the product of $k$ inner-products of 2-dimensional vectors $\bs{u}_{q}=(R_q, -iP_qR_q)^T$ and $\bs{a}(\theta)=(\cos{\theta},\sin{\theta})^T$
\begin{equation}
    U(\ptheta)=R_{p+1} (\bs{u}^T_{p} \bs{a}(\theta_p/2))\dots(\bs{u}^T_{1} \bs{a}(\theta_1/2)).
\end{equation}
This allows us to separate out the parameter-dependent part of the kernel from the ansatz-dependent part. We can then write the vectorized unitary $U(\ptheta)\otimes U^{*}(\ptheta)$ as 
\begin{equation}
    U(\ptheta)\otimes U^*(\ptheta)=R_{p+1}\otimes R_{p+1}^*(\bs{u}^T_{p} \bs{a}(\theta_p/2)\otimes (\bs{u}_{p}^*)^T \bs{a}(\theta_p/2))\dots(\bs{u}^T_{1} \bs{a}(\theta_1/2)\otimes (\bs{u}_{1}^*)^{T} \bs{a}(\theta_1/2)).
\end{equation}
The tensor products of the form $\bs{u}^T_{q} \bs{a}(\theta_q/2)\otimes (\bs{u}_{q}^*)^{T} \bs{a}(\theta_q/2)$ can be expanded and simplified into the following inner product
\begin{equation}
    \bs{u}^T_{q} \bs{a}(\theta_q/2)\otimes (\bs{u}_{q}^*)^{T} \bs{a}(\theta_q/2)=\frac{1}{2}\begin{pmatrix}
    (I \otimes I + P_q \otimes P_q^*)R_q\otimes R_q^*\\
    (I\otimes iP_q^* -iP_q\otimes I)R_q\otimes R_q^*\\
    (I \otimes I - P_q \otimes P_q^*) R_q\otimes R_q^* 
    \end{pmatrix}^T
    \begin{pmatrix}
    1\\
    \sin{\theta_q}\\
    \cos{\theta_q} 
    \end{pmatrix} = \bs{s}_q^T\bs{v}_q(\theta_q),
\end{equation}
where $\bs{v}_q=(1,\sin{\theta_q},\cos{\theta_q})^T$ and $\bs{s}_q$ is the vector of sums of unitaries above. The overall vectorized unitary is 
\begin{equation} \label{eq:UU}
    U(\ptheta)\otimes U^*(\ptheta)=R_{p+1}\otimes R_{p+1}^*(\bs{s}_p^T \bs{v}_p(\theta_p))\dots (\bs{s}_1^T \bs{v}_1(\theta_1))=R_{p+1}\otimes R_{p+1}^* \bs{s}^T\bs{v}(\bs{\theta}),
\end{equation}
where $\bs{s}$ is given by
\begin{equation}
    \bs{s}=\frac{1}{2^p}\left[
    \begin{pmatrix}
    (I \otimes I + P_p \otimes P_p^*)R_p\otimes R_p^*\\
    (I\otimes iP_p^* -iP_p\otimes I)R_p\otimes R_p^*\\
    (I \otimes I - P_p \otimes P_p^*) R_p\otimes R_p^* 
    \end{pmatrix}
    \otimes_K\dots\otimes_K
    \begin{pmatrix}
    (I \otimes I + P_1\otimes P_1^*)R_1\otimes R_1^*\\
    (I\otimes iP_1^* -iP_1\otimes I)R_1\otimes R_1^*\\
    (I\otimes I - P_1 \otimes P_1^*) R_1\otimes R_1^* 
    \end{pmatrix}
    \right]
\end{equation}
and $\bs{v}(\bs{\theta})$ by
\begin{equation}
    \bs{v}(\bs{\theta})=\begin{pmatrix}1\\\sin(\theta_p)\\\cos(\theta_p)\end{pmatrix}\otimes_K\dots\otimes_K\begin{pmatrix}1\\\sin(\theta_1)\\\cos(\theta_1)\end{pmatrix}.
\end{equation}
Here we have made a distinction between the tensor products $\otimes$ that denote operators on the physical and conjugated multi-qubit Hilbert spaces and the Kronecker products (for vectors with matrix entries) $\otimes_K$ between the $k$-``sub-vectors" which determine the dimension of the resulting kernels' feature spaces. The state kernel is then
\begin{equation}
\begin{split}
    k_s(\bs{\theta},\bs{\theta}') &= \bbra{\rho_0} \bs{v}^T(\bs{\theta}')\bs{s}^* \bs{s}^T \bs{v}(\bs{\theta})\kket{\rho_0}\\
    &= \bs{v}^T(\bs{\theta}')\bbra{\rho_0}\bs{s}^* \bs{s}^T\kket{\rho_0} \bs{v}(\bs{\theta})\\
    &= \bs{v}^T(\bs{\theta}')\bs{S} \bs{v}(\bs{\theta}),
\end{split}
\end{equation}
where each element of $\bs{s}^*$ is the Hermitian conjugate of the corresponding element in $\bs{s}$ so that $(\bs{S})_{ij}=\bbra{\rho_0}s_i^\dagger s_j\kket{\rho_0}$. (Note that the unitary $R_{p+1}\otimes R_{p+1}^*$ has been cancelled by its Hermitian conjugate.) 

The unitary kernel evaluated for this ansatz takes the following form:
\begin{equation}
\begin{split}
    k_u(\bs{\theta},\bs{\theta}')&=\frac{1}{d^2}\abs{\Tr{U^\dagger(\bs{\theta}') U(\bs{\theta})}}^2\\
    &=\frac{1}{d^2}\Tr{U^\dagger(\bs{\theta}') U(\bs{\theta})}\Tr{U^T(\bs{\theta}') U^*(\bs{\theta})}\\
    &=\frac{1}{d^2}\Tr{[U^\dagger(\bs{\theta}')\otimes U^T(\bs{\theta}')][ U(\bs{\theta})\otimes U^*(\bs{\theta})]}.
\end{split}
\end{equation}
Using the decomposition of this unitary given in \eqref{eq:UU}, we can write the unitary kernel as
\begin{equation}
\begin{split}
    k_u(\bs{\theta},\bs{\theta}')&=\frac{1}{d^2}\Tr{\bs{v}^T\bs{s}^*\bs{s}^T\bs{v}(\bs{\theta})}\\
    &=\bs{v}^T(\bs{\theta}')\frac{\Tr{\bs{s}^*\bs{s}^T}}{d^2}\bs{v}(\bs{\theta})\\
    &=\bs{v}^T(\bs{\theta}')\bs{T}\bs{v}(\bs{\theta}),
\end{split}
\end{equation}
where again each element of $(\bs{s}^*)_i=s_i^\dagger$ so that $\bs{T}={\Tr{\bs{s}^*\bs{s}^T}}/{d^2}$. (Again note that the $R_{p+1}$ terms have cancelled.)

Using the form for $U(\ptheta)\otimes U^*(\ptheta)$ given in \eqref{eq:UU} we also can now write the noiseless energy function $E(\ptheta)$ in terms of the $\bs{v}(\ptheta)$ vector, and thereby the feature space vectors, as 
\begin{equation}\label{eq:energy_linear_deriv}
\begin{split}
    E(\bs{\theta})&=\Tr{H U(\bs{\theta})\rho_0 U^\dagger(\bs{\theta})}=\bbra{H}U(\bs{\theta})\otimes U^*(\bs{\theta})\kket{\rho_0}\\
    &=\bbra{R_{p+1}^\dagger H R_{p+1}}\bs{s}^T\bs{v}(\ptheta)\kket{\vec(\rho_0)}\\
    &=\bbra{R_{p+1}^\dagger H R_{p+1}}\bs{s}^T\kket{\vec(\rho_0)}\bs{v}(\ptheta)\\
    &=\bs{h}^T\bs{v}(\bs{\theta}), \text{ where } \bs{h}=\bbra{R_{p+1}^\dagger H R_{p+1}}\bs{s}\kket{\vec(\rho_0)}.
\end{split}
\end{equation}

We can also use \eqref{eq:energy_linear_deriv} to calculate the mean value of the energy function $\mathbb{E}_{\ptheta}[E(\ptheta)]$ with respect to the gate angles, which is required to properly set the Gaussian process' mean. Because $E(\ptheta)$ is a linear function of $\bs{v}(\theta)$ we have
\begin{equation}
\mathbb{E}_{\ptheta}[E(\ptheta)] = \bs{h}^T\mathbb{E}_{\ptheta}[\bs{v}(\ptheta)].
\end{equation}
$\bs{v}(\ptheta)=\bs{v}^{(k)}\otimes_K\dots\otimes_K\bs{v}^{(1)}$ is a Kronecker product of $k$ subvectors with $\bs{v}^{(j)}=(1,\sin(\theta_j),\cos(\theta_j))^T$ each of which only depends on a single gate angle (we assume all gate angles are independent). This means that
\begin{equation}
\begin{split}
\mathbb{E}_{\ptheta}[\bs{v}(\ptheta)]=\frac{1}{{(2\pi)}^p}\int d^p\!\ptheta\ \bs{v}(\ptheta) &=\frac{1}{{(2\pi)}^p}\left(\int_0^{2\pi}\! \bs{v}^{(k)} d\theta_p\right)\otimes_K \ \dots\otimes_K \left(\int_0^{2\pi}\! \bs{v}^{(1)}d\theta_1 \right)\\
&=\begin{pmatrix}
1\\0\\0
\end{pmatrix}\otimes_K\dots\otimes_K \begin{pmatrix}
1\\0\\0
\end{pmatrix}.
\end{split}
\end{equation}
So the only the first element in $\mathbb{E}_{\ptheta}[\bs{v}(\ptheta)]$ is nonzero, greatly simplifying the expected energy to
\begin{equation}
    \mathbb{E}_{\ptheta}[E(\ptheta)] = h_0 = \bbra{R_{p+1}^\dagger H R_{p+1}}\frac{(R_p\otimes R_p^* + P_pR_p \otimes P_p^*R_p^*)}{2}\dots\frac{(R_1\otimes R_1^* + P_1R_1 \otimes P_1^*R_1^*)}{2}\kket{\vec(\rho_0)}.
\end{equation}
One way to interpret $\frac{1}{2}{(R_q\otimes R_q^* + P_qR_q \otimes P_q^*R_q^*)}$ is as a projector onto the positive eigenspace of the Pauli operator $P_q\otimes P_q^*$. Alternatively it can be seen as taking the part of its unvectorized input (following the application of $R_q$) that commutes with $P_q$. We can decompose an input $\rho'=R_q\rho R_q^\dagger$ into parts (sums of Pauli operators) that commute and anticommute with $P$, i.e.  $\rho'=c_q+a_q$ with $[P_q,c_q]=0$ and $\{P_q,a_q\}=0$ then $\frac{1}{2}{(R_q\otimes R_q^* + P_qR_q \otimes P_q^*R_q^*)}\kket{\rho} = \frac{1}{2}\kket{R_q\rho R_q^\dagger + P_qR_q\rho R_q^\dagger P_q}=\kket{c_q}$. Applying each of these projectors in this way we see that 
\begin{equation}
    \mathbb{E}_{\ptheta}[E(\ptheta)] = h_0 = \Tr{R^\dagger_{p+1}HR_{p+1}c_{1,\dots,k}}
\end{equation}
where $c_{1,\dots,k}$ is what remains of $\rho_0$ once the process of applying projector (applying each $R_q$ and removing the part of the result that anticommutes with $P_q$). In general calculating this is complicated as it depends on the fixed unitaries $\{R_q\}$ present in the ansatz however we suggest that for most ansatzes of interest the remainder will be $c_{1,\dots,k}=I/2^n$ as this factor is present in all density matrices (ensuring $\Tr{\rho}=1$) and commutes with any $P_q$ regardless of the fixed unitaries applied. This leads to our suggested approximation to the mean energy of $\mathbb{E}_{\ptheta}[E(\ptheta)] =h_0\approx\Tr{R_{p+1}^\dagger H R_{p+1}I/2^n}=\Tr{H}$. It takes only a few anti-commuting parameterized Pauli rotations on each qubit (e.g. two sequential non-commuting single-qubit PPRs on each qubit) for this approximation to be exact.

\section{Limits on state kernel feature space dimension}\label{appendix:fs_s}
In section \ref{sec:feat_construction} we identified the unnormalized states $\{s_i\kket{\rho_0}\}$ as vectorizations of 
Hermitian operators on $\mathcal{H}$. We can show this explicitly by considering the action of the $q^{\mathrm{th}}$ subvector of $\bs{s}$ \eqref{eq:s}, $\bs{s}^{(q)}$, on a vectorized input $\kket{\rho}$ (with Hermitian $\rho$):
\begin{equation}
    \bs{s}^{(q)}\kket{\rho}=\frac{1}{2}\begin{pmatrix}
    (I \otimes I + P_q \otimes P_q^*)R_q\otimes R_q^*\\
    (I\otimes iP_q^* -iP_q\otimes I)R_q\otimes R_q^*\\
    (I \otimes I - P_q \otimes P_q^*) R_q\otimes R_q^* 
    \end{pmatrix}\kket{\rho}=\frac{1}{2}\begin{pmatrix}
    \kket{R_q\rho R_q^\dagger + P_qR_q\rho R_q^\dagger P_q}\\
    \kket{iR_q\rho R_q^\dagger P - iP_qR_q\rho R_q^\dagger}\\
    \kket{R_q\rho R_q^\dagger - P_qR_q\rho R_q^\dagger P_q}
    \end{pmatrix}
\end{equation}
As in the discussion given in given in Appendix \ref{appendix:feature_spaces} on the mean energy, if we write $\rho'=R_q\rho R_q^\dagger$ as a sum of a commuting and an anti-commuting part $P$, i.e.  $\rho'=c_q+a_q$ with $[P_q,c_q]=0$ and $\{P_q,a_q\}=0$, then we can rewrite the resulting vector as
\begin{equation}\label{eq:subvector_action}
    \bs{s}^{(q)}\kket{\rho}=\begin{pmatrix}
    \kket{c_q}\\
    \kket{ia_qP_q}\\
    \kket{a_q}
    \end{pmatrix}.
\end{equation}
So the $q^\mathrm{th}$ subvector of $\bs{s}$ acts to separate the input $\rho'$ (transformed by $R_q$) into the parts which commute (first entry) and anti-commute (last entry) with $P_q$ while the middle entry contains the anti-commuting part post-multiplied by $iP_q$. Like the input, all of these operators are Hermitian. The first and last entries are formed by separating Pauli operators in the decomposition of the input into commuting and anti-commuting subsets; this means that if $\rho'$ has support on (can be decomposed into) a finite set of $m$ Pauli operators, $\rho'=\sum_{i=1}^m P_i$, then Pauli operators that do not feature in this decomposition can only be introduced in the middle entry. We also note that $c_q$, $ia_qP_q$, and $a_q$ are necessarily linearly independent as it can be shown using the cylic property of the trace, $PP=I$, and $P_qa_q=-a_qP_q$, that they are mututally orthogonal under the Hilbert-Schmidt norm $\langle A,B\rangle_{HS}=\Tr{A^\dagger B}$; $\langle a_q,c_q\rangle_{HS}=-\langle a_q,c_q\rangle_{HS}=0$, $\langle ia_qP_q,c_q\rangle_{HS}=-\langle ia_qP_q,c_q\rangle_{HS}=0$, $\langle ia_qP_q,a_q\rangle_{HS}=-\langle ia_qP_q,a_q\rangle_{HS}=0$.

The dimension of the feature space is determined by the number of parameterized rotations (increasing the dimension of $\bs{v}(\ptheta)$) and the rank of the Gram matrix $\bs{S}$. Each unnormalized state $s_i\kket{\rho_0}=\kket{O_i}$ is a vectorization of a Hermitian operator $O_i$ meaning the inner products forming $S_{ij}=\bbra{\rho_0} s_i^\dagger s_j\kket{\rho_0}$ can be written $S_{ij}=\Tr{O_iO_j}$, the Hilbert-Schmidt inner-product between $O_i$ and $O_j$. To maximise the dimension of $\mathcal{F}_s$ we must ensure that the $\{O_i\}$ for different entries of $\bs{s}\kket{\rho_0}$ are linearly independent (spanning a subspace of Hermitian operators of dimension $\geq 3^p$, the number of entries). 

Suppose after the $j^\mathrm{th}$ subvector has been applied, the vector $s^{(j)}\otimes_K\dots\otimes_K s^{(1)}\kket{\rho_0}$ has $m$ linearly independent elements/operators. We can write each element as a sum of an operator that commutes with the next parameterized Pauli rotation $P_{j+1}$ and one that anti-commutes: $s^{(j)}\otimes_K\dots\otimes_K s^{(1)}\kket{\rho_0}=(\kket{a_1+c_1},\dots,\kket{a_m+c_m})^T$. Applying the subvector for $P_{j+1}$ splits each element into $3$ new elements, e.g.  $\kket{a_1+c_1}\to(\kket{c_1},\kket{ia_1P_{j+1}},\kket{a_1})$, increasing the feature space dimension by a factor of $3$. However, these elements will not always be linearly independent. Clearly, the contribution to the number of linearly independent elements from the commuting terms $c_1\dots c_m$ is at most of size $m$. From the anti-commuting terms we separate out introduce the elements $a_1,\dots,a_m$ and $ia_1P_{j+1},\dots,ia_mP_{j+1}$, which also anti-commute with $P_{j+1}$. An $n$-qubit Pauli operator anti-commutes with $4^{n}/2$ other elements of the $n$-qubit Pauli group (ignoring phases) and commutes with the other remaining  $4^{n}/2$ elements. As there are only $4^n/2$ Pauli operators that anti-commute with $P_{j+1}$, the span of the combined set of $a_1,\dots, a_m, ia_1P_{j+1},\dots, ia_mP_{j+1}$ is at most $4^n/2$ dimensional. This means that the maximal increase in the feature space dimension possible when adding another parameterized Pauli rotation to the circuit is $4^n/2$ (as we can go from $m$ independent elements to at most $m+4^n/2$). Denoting the maximal feature space dimension of $\mathcal{F}_s$ for $n$ qubits and a $p$ parameterized Pauli rotation ansatz as $d_s^{(n,p)}$ we have the recursion relation

\begin{equation}
    d_s^{(n,p)}=\min(4^n, 3d_s^{(n,p-1)}, 4^n/2 + d_s^{(n,p-1)}).
\end{equation}
We define the base-case $d_s^{n,0}=1$ for all $n$ -- this is purely for completeness as a kernel on a zero-dimensional input (no parameterized rotations) has no meaning.

\section{Limits on unitary kernel feature space dimension}\label{appendix:fs_u}
Like the state kernel, the unitary kernel's feature space can be expressed as a quadratic form involving a parameter-dependent vector $\bs{v}(\ptheta)$ and a Gram matrix $\bs{T}$ with elements $T_{ij}=\Tr{s_i^\dagger s_j}$. Like with the state kernel, the dimension of the feature space $\mathcal{F}_u$ is determined by the number of parameters (determines the size of $\bs{v}$) and the rank of the $\bs{T}$ matrix. 

The $T$ matrix is comprised of Hilbert-Schmidt inner products between the elements of the vector $\bs{s}$. If the elements of $\bs{s}$ were completely generic $2^{2n}\times2^{2n}$ matrices then the maximal rank of $\bs{T}$ would be $4^{2n}$ as there are this many orthogonal (under the Hilbert-Schmidt inner product) basis matrices, e.g. the $2n$-qubit Pauli matrices $\{I,X,Y,Z\}^{\otimes 2n}$ or the standard basis matrices $e_{ij}=\ketbra{i}{j}$. However there is additional structure present in the elements of $\bs{s}$ that reduces the maximal rank of $\bs{T}$. For example, all elements in $\bs{s}$ are symmetric under interchange of the physical ($\mathcal{H}$) and conjugated ($\mathcal{H}^*$) subsystems followed by complex conjugation and involve sums and products of various unitary and Hermitian operators. To derive the maximal scaling of the unitary kernel's feature space dimension we will first examine weighted sums of unitary matrices of the form $A\otimes A^*$, matrices $\bs{M}(\bs{w}, \bs{A})$ given by 
\begin{equation}\label{eq:w_sum}
    \bs{M}(\bs{w}, \bs{A})=\sum_k w_k A^{(k)}\otimes {A^{(k)}}^*
\end{equation}
with ${A^{(k)}}^\dagger A^{(k)} = I_d$ (and each $A$ has dimension $d\times d$). In the following we will demonstrate that the maximal dimension of $\mathcal{F}_u$ is ultimately set by the number of linearly independent entries of such matrices.

For a single $A\otimes A^*$, the unitarity constraint implies that $\sum_k A_{ki} A^*_{kj} = \sum_k A_{ik} A^*_{jk}  = \delta_{ij}$ -- these constraints can be easily related to $A\otimes A^*$ as they correspond to contractions over pairs of indices. A general element of this matrix is $(A\otimes A^*)_{ijkl}=A_{ij}A^*_{kl}$ so contraction over the first and third indices gives $\sum_p (A\otimes A^*)_{kikj} = \sum_p A_{ki} A^*_{kj} = \delta_{ij}$ (the first set of unitarity constraints) while contraction over the second and fourth indices gives $\sum_p (A\otimes A^*)_{ikjk} = \sum_p A_{ik} A^*_{jk} = \delta_{ij}$ (the second set of constraints). This appears to give us $2d^2$ equations, each of which will reduce the number of linearly independent elements by $1$ giving a total of $d^4-2d^2$ independent elements. However, because the constraints for $i=j$ all sum to the same value, $\sum_p A_{ik}A^*_{ik}=\sum_p A_{ki} A^*_{ki} = 1$ for all $i$, this fixed value acts as another free parameter in each set of constraints, increasing the number of linearly independent elements by $2$ to a total of $d^4-2(d^2-1)$. 
When considering weighted sums of these matrices as given in \eqref{eq:w_sum} these constraints become
\begin{equation}\label{eq:uni_constraint}
    \sum_p \sum_p (w_p A^{(p)}\otimes {A^{(p)}}^*)_{ikjk} = \sum_p \sum_p (w_p A^{(p)}\otimes {A^{(p)}}^*)_{kikj} = (\sum_p w_p)\delta_{ij}
\end{equation}

The linear dependence of the terms in the constraints with $i=j$ is only defined up to specification of the sum of the weights meaning that the quantity $\sum_p w_p$ is an additional free parameter, giving the same result.  

The initial fixed unitary $R_1\otimes R_1^*$ is in the form we have just discussed and for an $n$-qubit ansatz will in general have $d^4-2(d^2-1)=4^{2n}-2(4^n-1)$ linearly independent elements. For unitary operators $B$ and $C$, the product $(B\otimes B^*)(C\otimes C^*)=A\otimes A^*$ (with $A=BC$) is also a matrix of the form discussed above. As a result, the Pauli-containing part of the top/bottom subvector elements $I\otimes I \pm P_1\otimes P_1^*$ acts on $R_1\otimes R_1^*$ to give matrices that admit a decomposition as in \eqref{eq:w_sum} and so have the same number of linearly independent elements as before (as will the result from applying the next fixed unitary $R_2\otimes R_2^*$ and the subsequent top/bottom elements of the next subvector). All that is left is to consider how the application of $i(I\otimes P^*-P\otimes I)$, from the middle elements of each subvector, to a general unitary $A\otimes A^*$ (which could be part of a sum as in \eqref{eq:w_sum}) affects the unitarity constraints \eqref{eq:uni_constraint} that determine the number of linearly independent elements. We have:
\begin{equation}
\begin{split}
 \sum_k (i(I\otimes P^*-P\otimes I)(A\otimes A^*))_{ikjk} &= i\sum_p \sum_m( {A}_{ik}P^*_{jm}A^*_{mk} -P_{im}A_{mk}A^*_{jk})\\
 &=i\sum_m( \delta_{im}P^*_{jm} - P_{im}\delta_{mj}) = i(P^*_{ji} - P_{ij})\\
 &=i(P^\dagger-P)_{ij} = 0,
\end{split}
\end{equation}
while for the other constraint we get the same result
\begin{equation}
\begin{split}
 \sum_p (i(I\otimes P^*-P\otimes I)(A\otimes A^*))_{kikj} &= i\sum_p \sum_m( {A}_{ki}P^*_{km}A^*_{mj} -P_{km}A_{mi}A^*_{kj})\\
 &=i\sum_p \sum_m( P_{mk}{A}_{ki}A^*_{mj}) -i\sum_m \sum_p(P_{km}A_{mi}A^*_{kj})\\
 &= 0
\end{split}
\end{equation}
(in both cases we have used that $P^\dagger = P$). By linearity, the same results will be obtained for an arbitrary weighted sum as in \eqref{eq:w_sum}. These are identical to the constraints in \eqref{eq:uni_constraint} for $i\neq j $ but give $0$ rather than a fixed value ($1$ for a single $A\otimes A$ or $\sum_p w_p$ for a weighted sum) for $i=j$. Having the $i=j$ case evaluate to $0$ means that the number of linearly independent elements in matrices formed of weighted sums of terms like this is $d^4 - 2(d^2)=4^{2n}-2(4^n)$. However, because these new constraints are a subset of the original constraints \eqref{eq:uni_constraint} with $\sum_p w_p=0$, the over-all number of linearly independent elements in the matrices that form the entries of $\bs{s}$ and therefore the maximal rank of $\bs{T}$ is given by $4^{2n}-2(4^n-1)$. We verified this scaling with simulations for $n=1,2, \text{ and } 3$ qubits using Haar random unitaries for $\{R_q\}$ to ensure minimal structure in the ansatz that might limit the feature space dimension.

\end{document}